\documentclass[10pt,journal]{IEEEtran}
\IEEEoverridecommandlockouts

\usepackage{graphicx}
\usepackage{subfigure}
\usepackage{amsmath,amsthm,amsfonts,amssymb}
\usepackage{bm}
\usepackage{bbm}
\usepackage{url}
\usepackage{array}
\usepackage{color}
\usepackage{multirow}
\usepackage{booktabs}
\usepackage[table,xcdraw]{xcolor}
\usepackage{enumerate}
\usepackage{cite}

\theoremstyle{plain}
\newtheorem{thm}{Theorem}
\newtheorem{lem}[thm]{Lemma}

\newtheorem{prop}[thm]{Proposition}
\newtheorem{cor}[thm]{Corollary}

\newtheorem*{rem}{Remark}
\newtheorem{sty1}{Theorem}
\newtheorem{defi}[sty1]{Definition}

\newenvironment{NewProof}{{\noindent\it Proof.}}{\hfill $\blacksquare$\par}

\usepackage[english]{babel}
\usepackage{algorithm}
\usepackage[noend]{algpseudocode}

\allowdisplaybreaks[4]

\begin{document}
\title{Partially Observable Minimum-Age Scheduling:\\ The Greedy Policy}

\author{Yulin~Shao,~\IEEEmembership{Member,~IEEE},
Qi~Cao,
Soung~Chang~Liew,~\IEEEmembership{Fellow,~IEEE},
He~Chen,~\IEEEmembership{Member,~IEEE}
\thanks{The authors are with the Department of Information Engineering, The Chinese University of Hong Kong, Shatin, New Territories, Hong Kong (e-mail: \{ylshao, cq016, soung, he.chen\}@ie.cuhk.edu.hk).}
}

\maketitle

\begin{abstract}
This paper studies the minimum-age scheduling problem in a wireless sensor network where an access point (AP) monitors the state of an object via a set of sensors. The freshness of the sensed state, measured by the age-of-information (AoI), varies at different sensors and is not directly observable to the AP. The AP has to decide which sensor to query/sample in order to get the most updated state information of the object (i.e., the state information with the minimum AoI). In this paper, we formulate the minimum-age scheduling problem as a multi-armed bandit problem with partially observable arms and explore the greedy policy to minimize the expected AoI sampled over an infinite horizon. To analyze the performance of the greedy policy, we 1) put forth a relaxed greedy policy that decouples the sampling processes of the arms, 2) formulate the sampling process of each arm as a partially observable Markov decision process (POMDP), and 3) derive the average sampled AoI under the relaxed greedy policy as a sum of the average AoI sampled from individual arms. Numerical and simulation results validate that the relaxed greedy policy is an excellent approximation to the greedy policy in terms of the expected AoI sampled over an infinite horizon.
\end{abstract}


\section{Introduction}\label{sec:introduction}
Information freshness has long been an important Quality of Service (QoS) consideration in communication networks \cite{AoI1,AoI2,Significant2020,AoI3,techFL}. In the 5G era, a host of mission-critical applications, e.g., robotic control in Industrial Internet of Things (IIoT) \cite{IIoT}, vehicle-to-vehicle communications (V2V) \cite{V2X}, requires sub-millisecond end-to-end latency to guarantee the prompt delivery of the time-critical information.

The {\it Age-of-Information} (AoI), originally proposed in \cite{AoI1,AoI2}, is a new performance metric capturing information freshness from the receiver's perspective.
Specifically, AoI measures the time elapsed since the generation of the freshest packet delivered to the receiver. In comparison, the traditional {\it latency} metric measures the time consumed by queuing and transmission/propagation from the transmitter's perspective.
The advent of AoI has shed new light on the design and optimization of wireless networks to support time-sensitive applications.

\subsection{Minimum-age Scheduling in Sensor Networks}
\begin{figure}[t]
  \centering
  \includegraphics[width=0.6\columnwidth]{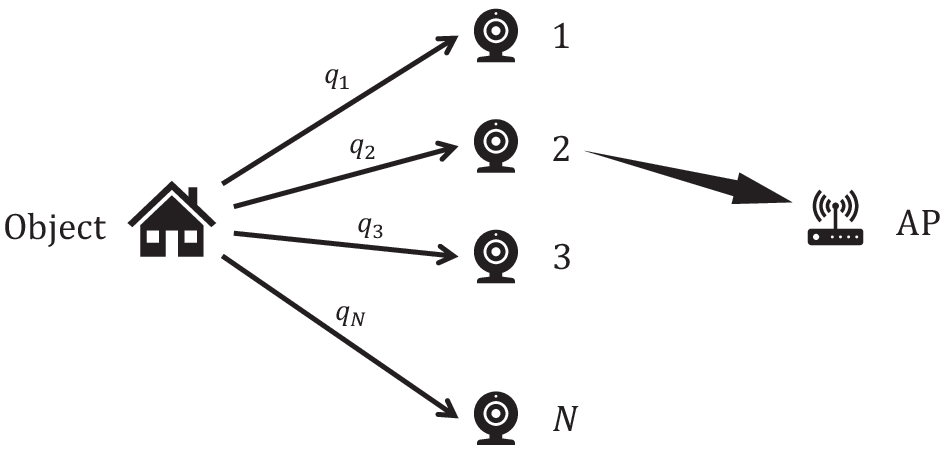}\\
  \caption{A wireless sensor network where the AP monitors the state of an object through a set of randomly deployed sensors. The sensing channels are unreliable and mutually independent. In each slot, only one sensor can report its sensed information to the AP.}
\label{fig:model}
\end{figure}

This paper considers a general minimum-age scheduling problem in wireless sensor networks. The system model is shown in Fig.~\ref{fig:model}, where an access point (AP) monitors the state of an object or process via multiple randomly-deployed sensors. The sensing channels are unreliable and mutually independent. As a result, the ages of the sensed states at different sensors vary.

We assume a time-slotted model. The age of the sensed state (i.e., AoI) at each sensor is updated to one if the sensor successfully receives the state of the object at the end of a time slot. Otherwise, the AoI increases by one. The dynamic of the aging process can thus be captured by a Markov chain with a countably infinite number of states, wherein the state of the Markov chain corresponds to the AoI of the sensed state of the object.

In each time slot, the AP queries/samples one of the sensors to collect its sensed state. However, the real-time AoIs of all sensors are unknown to the AP because transmitting AoI information consumes extra energy of the sensors. Therefore, the AP can obtain the AoI of a sensor only when the AP samples it. The minimum-age scheduling problem considered in this paper is as follows: {\it at any slot, given a sequence of past sampling decisions and observations, which sensor should the AP sample to minimize the expected AoI sampled over an infinite horizon}?

Mathematically, the problem under study belongs to a class of Multi-Armed Bandit (MAB) problems \cite{GittinsBook}. The MAB problem is a stochastic control problem wherein a controller sequentially allocates a limited resource amongst alternative arms so as to minimize the costs incurred by the allocations. In our case, the arms are the sensors, the limited resource is the channel access opportunity in each slot for a sensor to report its sensed information to the AP, and the cost is the expected AoI sampled from the sensors over an infinite horizon.

In classical MAB, only the chosen arm can evolve and incur costs, while the unchosen arms remain frozen \cite{GittinsBook,weber1992gittins}. This framework was then generalized by Whittle \cite{Whittle1988} to a restless MAB (RMAB) wherein the state of each arm evolves continuously, whether it is sampled or not. In Whittle's original formulation, the states of the arms are fully observable to the controller, whereas in our problem, the AoIs of the sensors are partially observable. Thus, our problem is a partially observable RMAB.

The considered system model has a variety of applications in practice. Due to the unreliable nature of randomly-deployed sensors, often a set of sensors (instead of one) are deployed in the area of interest so that a subset of sensors can be in good positions to monitor the target. Typical applications include underwater sensing \cite{r1}, smart city \cite{r2}, military surveillance \cite{r3}, intelligent agriculture \cite{r4}, and planetary exploration \cite{r5}, especially when the area of interest is harsh, hostile, or inaccessible via conventional means \cite{r6}. Two specific examples are given as follows:
\begin{enumerate}
\item Environmental monitoring. To monitor a forest fire or a volcanic eruption, sensors are air-dropped from an unmanned aerial vehicle (UAV) and randomly placed in the target area. Different sensors observe the environment (or the target) independently and the AoI of the sensed data varies. To obtain the freshest information, the UAV samples the sensors by periodically broadcasting a beacon signal and asking one of the sensors to report its sensed data \cite{r6,PNC}.
\item Military surveillance. On the battlefield, sensors are randomly deployed in remote areas to monitor the militant activity; acoustic sensor arrays are deployed underwater to detect transient signals from mortars, artillery and small arms fire \cite{r3,r6}. An AP samples the randomly deployed sensors to collect the freshest information, following the system model considered in this paper.
\end{enumerate}

\subsection{Related Work}
\textbf{Minimum-age scheduling} -- Much of the AoI research efforts have been devoted to the minimum-age scheduling problem in centralized networks where a central controller coordinates all the transmissions in the network \cite{Kadota1,Kadota2,MIT3,generateatwill,flow,AoI_queue,LCFS}. The research focus is on the scheduling policies of the coordinator to minimize the expected weighted sum of the AoIs of all users in the network. Various network architectures (e.g., broadcast \cite{Kadota1,Kadota2}, multiple access \cite{MIT3}), traffic arrival models (e.g., generate at will \cite{generateatwill,flow}, stochastic arrivals \cite{Kadota2,randomarrival}), and queueing models (e.g., M/G/1 \cite{AoI_queue}, last-come-first-served \cite{LCFS}) have been considered. The readers are referred to a recent survey paper \cite{survey1} for a detailed treatment of the state-of-the-art in this domain.

The difference between prior works and our study in this paper is the observability of the AoI at the users/sensors. In prior arts, the AoI of the users are assumed to be fully known to the central controller at any time, whereby the scheduling decisions can be made based on the exact AoI of users. In our problem, however, the AoIs of users at a decision epoch are partially observable -- they can only be inferred from past decisions and observations. This leads to a fundamental difference in the analytical approaches.

\textbf{RMAB} -- RMAB is closely related to our problem. One standard way to solve RMAB is to use the value iteration algorithm of the Markov decision process (MDP) theory \cite{MDPBook} since the MAB problems themselves are MDP problems. However, the complexity of value iteration grows exponentially with the number of arms.

On the other hand, as shown by Gittins \cite{GittinsBook} and Whittle \cite{Whittle1988}, MAB problems admit index policies, the complexity of which only increases linearly with the number of arms. The heuristic index policy proposed by Whittle to solve RMAB is known as the Whittle index policy \cite{Whittle1988}. The basic idea is to decouple the RMAB problem to multiple single-armed-bandit problems (by a Lagrangian relaxation) so that the arms are independent of each other. After decoupling, the scheduling problem associated with a single arm is then modeled as single-bandit MDP, whereby a Whittle index is computed. Given the Whittle indexes computed for individual arms, the controller simply chooses the arm with the largest index in each decision epoch. For this approach to be viable, however, the single-bandit problem must have the indexability property \cite{GittinsBook}.

Whittle's original formulation assumed fully observable arms, but it is possible to apply the Whittle index to partially observable arms \cite{RMABUAV,RMABSensor}, in which case the scheduling problem associated with a single arm is modeled as a partially observable MDP (POMDP) \cite{SondikThesis,POMDPBook}, as opposed to the MDP in the fully-observable case. However, a problem is that POMDPs are polynomial space (PSPACE) hard to solve, the optimal policy of which is tractable only when strict assumptions are made on the POMDP model \cite{monotone3}.

Existing works on the Whittle index's approach to solve the partially observable RMAB \cite{RMABUAV,RMAB2statePOMDP1,RMAB2statePOMDP2,RMAB2statePOMDP3,RMAB2statePOMDP4,RMAB2statePOMDP5,Gongpu}, to the best of our knowledge, are limited to the case where the state of each arm evolves as a two-state (on-off) Markov chain. Ref \cite{RMAB2statePOMDP1}, for example, proved the indexability of binary RMABs and derived the Whittle index in closed-form considering a linear cost function. Our recent work, \cite{Gongpu}, further generalized the Whittle index approach to  RMABs with any concave cost functions, provided that the underlying Markov chains are binary.
Yet, the general partially observable RMAB problem beyond the two-state arms, as the problem faced by this paper, is still open.

\textbf{POMDP} -- POMDP is a useful framework to model sequential decision problems with incomplete state information. To minimize the long-term average cost, an agent performs actions in the environment based on its observations on the system states. In particular, the observations contain only partial information of the current system state, from which the agent forms a belief (in the form of a probability distribution) on the current state the system. The execution of an action steers the environment to a new state and incurs a cost to the agent. The optimal solution to the POMDP is then the policy that yields the minimum cost over an infinite horizon at each decision epoch. In this context, the partially observable RMAB problems are POMDPs.

POMDP problems are PSPACE hard as they require exponential computational complexity and memory \cite{POMDPBook}. The optimal policy to a POMDP is analyzable only when strict assumptions are made on the POMDP model. A series of notable works that developed the structural results for the optimal policy of POMDPs can be found in \cite{monotone1,monotone2,monotone3}. Specifically, the authors aim to establish sufficient conditions on the cost function, dynamics of the Markov chain, and observation probabilities so that the optimal policy to the POMDP presents a threshold structure with respect to a monotone likelihood ratio (MLR) ordering. By doing so, the computational complexity of the optimal policy is inexpensive.

For general POMDP models that do not satisfy the sufficient conditions, investigators resorted to heuristic policies (e.g., the index policy \cite{Whittle1988,RMAB2statePOMDP1}, the greedy policy \cite{myopic1,myopic2,myopic3}), or suboptimal algorithms (e.g., Lovejoy's algorithm \cite{suboptimalAlgo_lovejoy}, point-based methods \cite{suboptimalAlgo_pointbased}). As will be shown later, the problem considered in this paper does not satisfy the sufficient conditions established in the above mentioned works, hence the monotonicity of the optimal policy is unknown.

\subsection{Contributions}
In the context of minimum-age scheduling, this paper studies the partially observable RMAB problem where each arm has a large number of states. The optimal policy to this problem, which minimizes the long-term expected AoI, is not practically computable due to the PSPACE hardness of POMDP. In this paper, we explore the greedy policy that minimizes the immediate expected AoI to solve this problem. Despite the simple descriptions, the greedy policy is by no means trivial to analyze since the AP observes a POMDP governed by multiple non-binary Markov chains -- this leads to two main obstacles:
\begin{enumerate}[a)]
\item The POMDPs associated with individual sensors are coupled together since the evolutions of the POMDPs are steered by the same sampling decision.
\item The sampling decisions over consecutive time slots are coupled together because a decision steers the POMDP to a new state, from which a later decision is made.
\end{enumerate}

To circumvent the performance-analytical challenge a), we put forth a relaxed greedy policy as an approximation to the greedy policy. Unlike the greedy policy, the relaxed greedy policy allows the AP to sample all the sensors whose expected AoI is less than a constant. The constant is carefully chosen so that the AP samples on average one sensor per time slot. In so doing, the sampling processes (and hence the POMDPs) associated with individual sensors are decoupled, the AoI sampled per slot is then the sum of the expected AoI sampled from each sensor.

For the decoupled POMDP, we tackle challenge b) by exploiting the recurrent structure of the POMDP. The expected AoI sampled from each sensor is found to satisfy a set of recurrence relations and can be derived by finding a particular solution to the recurrence relations.
Numerical and simulation results validate that the relaxed greedy policy is an excellent approximation to the greedy policy as far as the expected AoI sampled over an infinite horizon is concerned.

\section{System Model}\label{sec:II}
We consider a wireless sensor network wherein $N$ sensors are randomly deployed in an area to monitor the state of an object or process, as shown in Fig.~\ref{fig:model}. An access point (AP) collects the sensed data from the $N$ sensors by periodically broadcasting a beacon signal and asking one of the sensors to report its sensed data. As such, the sensors are aligned by the beacon signal in a time-slotted manner.

\subsection{Age of Information of the Sensors}
In a time slot, the probability that $n$-th sensor successfully captures the state of the object is $q_n$. Thus, if we denote by $\{H_n^t:n=1,2,...,N\}$ the event that the $n$-th sensor captures the state of the object in slot $t$, then $H_n^t$ follows Bernoulli distribution with parameter $q_n$, and is time-invariant (constant over time).

The $N$ sensors monitor the same information, i.e., the state of the object, but their information freshness can be different owing to the probabilistic sensing channels.\footnote{More generally, the problem being considered in this paper is relevant to a scenario in which a monitor attempts to monitor the state of an entity. The state of entity is collected by a set of state collector. The state information possessed by the collectors has varying degrees of out-datedness because of the varying random delays and reliabilities in their state collection processes. The monitor has to decide which collector to query in order to get the most updated state information of the entity.} The information freshness of a sensor is measured by the age-of-information (AoI).

\begin{defi}[AoI of sensors]
The Age of Information of the $n$-th sensor at the end of time slot $t$, denoted by $a_n^t$, is the number of slots elapsed since the last slot the $n$-th sensor successfully sensed the state of the object. Specifically, if a sensor successfully sensed the data in slot $t$, the AoI of this sensor at the end of slot $t$ is updated to $a_n^t=1$ (the sensed data is already one slot old); if the sensor failed to sense the data in slot $t$, the AoI of this sensor at the end of slot $t$ is increased by $1$, i.e., $a_n^t=a_n^{t-1}+1$. That is,
\begin{eqnarray}
a_n^t = \begin{cases}
1& \textup{w.~p.}~q_n,\\
a_n^{t-1}+1& \textup{w.~p.}~p_n=1-q_n.
\end{cases}
\end{eqnarray}
\end{defi}

\begin{figure}[t]
  \centering
  \includegraphics[width=0.7\columnwidth]{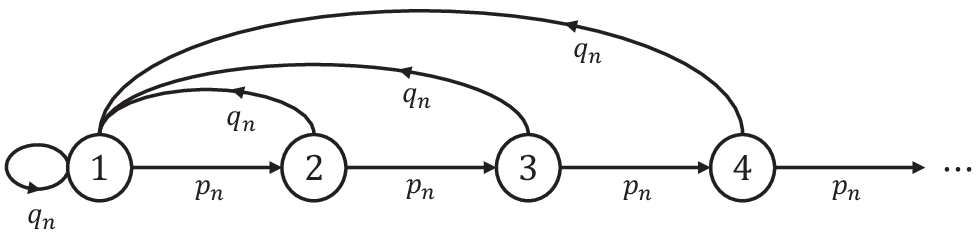}\\
  \caption{The AoI transitions of the $n$-th sensor as a discrete-time MC.}
\label{fig:MC}
\end{figure}

Let us define the AoI of a sensor as its {\it state}. As illustrated in Fig.~\ref{fig:MC}, the state transitions of each sensor form a discrete-time Markov chain (MC). The transition matrix of this MC is given by
\begin{eqnarray}
\bm{\mathcal{T}}_n =
\begin{bmatrix}
q_n & p_n & 0 & 0 & \cdots \\
q_n & 0 & p_n & 0 & \cdots \\
q_n & 0 & 0   & p_n  & \cdots \\
q_n & 0 & 0 & 0 & \cdots \\
\vdots & \vdots & \vdots & \vdots & \ddots
\end{bmatrix}.
\end{eqnarray}

The Markov chain shown in Fig.~\ref{fig:MC} has an infinite number of states. To ease analysis, we consider a truncated version of the MC with a finite number of $M$ states. That is, AoI equals or larger than $M$ are grouped as a single state ($M$ can be very large to avoid the impact of AoI truncation). After truncation, the $M\times M$ transition matrix becomes
\begin{eqnarray}\label{eq:matrix}
\bm{\mathcal{T}}_n = \begin{bmatrix}
q_n    & p_n    & 0      & \cdots & 0      & 0 \\
q_n    & 0      & p_n    & \cdots & 0      & 0\\
q_n    & 0      & 0      & \ddots & 0      & 0\\
\vdots & \vdots & \vdots & \vdots & p_n    & 0\\
q_n    & 0      & 0      & \cdots      & 0      & p_n\\
q_n    & 0      & 0      & \cdots      & 0      & p_n
\end{bmatrix}.
\end{eqnarray}

\subsection{Sampling Policy of the AP}
Now that the AoIs of different sensors vary, the AP aims to query/sample the sensor with the minimum AoI in each slot as it carries the freshest state information of the object.
The sampling process in each slot is error-free thanks to the error correction code in the PHY layer and the automatic repeat request (ARQ) protocol in the MAC layer.

\begin{figure}[t]
  \centering
  \includegraphics[width=0.92\columnwidth]{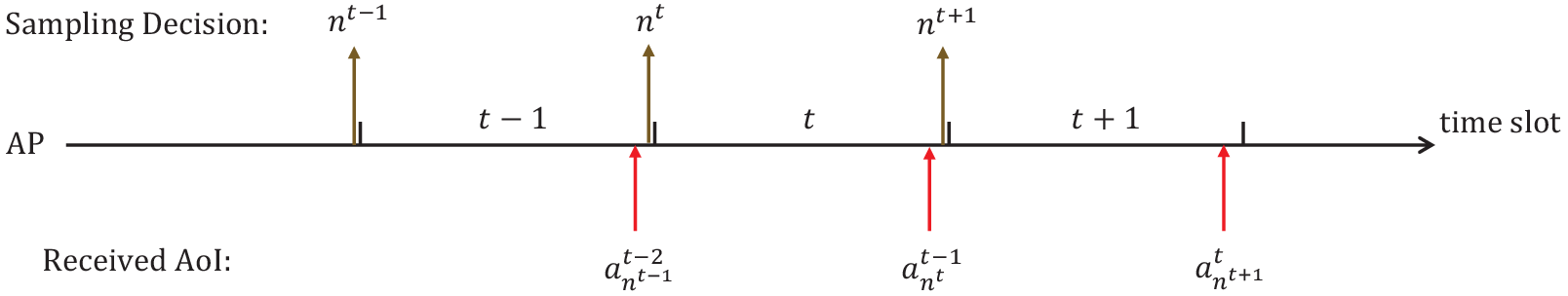}\\
  \caption{The sampling process and the AoI update from the sensors to the AP in consecutive time slots. A sampling decision is made before a slot begins, and the sensed data about the object is delivered to the AP by the end of a slot.}
\label{fig:timeline}
\end{figure}

Fig.~\ref{fig:timeline} shows the sampling process and the AoI update from the sensors to the AP in consecutive time slots.
At the beginning of a slot $t$, the AP broadcasts the beacon signal and query/sample one of the sensors, say, the $n^t$-th sensor. The sampling decision, i.e., which sensor to sample in slot $t$, is made at the end of the time slot $t-1$. Then, at the end of the slot $t$, the AP receives the feedback from the $n^t$-th sensor and the age of the received information is $a_{n^t}^{t-1}$, i.e., the AoI of the $n^t$-th sensor at the end of slot $t-1$ (the received AoI is $a_{n^t}^{t-1}$ rather than $a_{n^t}^{t}$ because the $n^t$-th sensor transmits the sensed data upon receiving the beacon signal at the beginning of the slot $t$, at which time $a_{n^t}^t$ is unknown). Next, the AP has to determine which sensor to sample in the next slot $t+1$, and the cycle continues.

When making the sampling decision at the end of the slot $t$, the exact AoIs of sensors $\{a_n^t:\allowbreak n=1,2,...,N\}$ are not directly observable by the AP. The only information known to the AP that can be used to make the sampling decision is a sequence of past decisions and observations $\mathcal{Z}^t=\{n^{t-\tau},\allowbreak a_{n^{t-\tau}}^{t-\tau-1}:\allowbreak\tau=0,1,2,3,...\}$, where $n^{t-\tau}$ and $a_{n^{t-\tau}}^{t-\tau-1}$ are the indexes of the sampled sensors in slot $t-\tau$ and the observed AoI at the end of slot $t-\tau$, respectively.

Given $\mathcal{Z}^t$, a sampling policy $\mu$ makes the sampling decision by $n^{t+1}=\mu(\mathcal{Z}^t)$, and receives an AoI $a_{n^{t+1}}^t$ at the end of slot $t+1$. Over time, given a sequence of sampled AoI $\{a_{n^t}^{t-1}:\allowbreak t=1,2,3,...\}$, the performance of the sampling policy $\mu$ is measured by the expected sampled AoI over the infinite horizon
\begin{eqnarray}\label{eq:metric}
J(\mu)=\mathbb{E}_\mu\left[\lim_{T\to\infty}\frac{1}{T}\sum_{t=0}^{T-1} a_{n^t}^t \right].
\end{eqnarray}
The optimal sampling policy, $\mu^*$, is the policy that minimizes $J(\mu)$, giving,
\begin{eqnarray}\label{eq:optpolicy}
\mu^*=\arg\min_\mu J(\mu).
\end{eqnarray}

\section{A POMDP formulation}\label{sec:III}
The decision problem in \eqref{eq:optpolicy} is essentially a POMDP: the system dynamics are governed by the $N$ Markov Chains of the $N$ sensors, the real-time state of which are unknown to the AP. At a decision epoch, the AP determines which sensor to sample based on a set of past decisions and observations $\mathcal{Z}^t$ because the instantaneous states (AoI) of the sensors are unobservable.
\subsection{Belief-state POMDP}
To the AP, the instantaneous state of each sensor at the end of slot $t$ is a random variable, denoted by $A_n^t$, distributed on $\{a_n^t=1,2,\cdots,M\}$. Although the exact states of the sensors are unknown, a joint distribution of the random variables $\{A_n^t:n=1,2,\cdots,N\}$ can be constructed from the history $\mathcal{Z}^t$.

\begin{defi}[belief state of the $N$ sensors]
The belief state of the POMDP at the end of slot $t$ is an $MN$-dimensional posterior probability distribution
\begin{equation*}
\bm{\Pi}^t =\left\{\Pi^t(a_1, a_2, ...,a_N): a_1,a_2,..., a_N\in\{1,2,...,M\}\right\},
\end{equation*}
where each entry
$\Pi^t(a_1,a_2,...,a_N)=\allowbreak\textup{Pr}(A^t_1=a_1\!\mid\!\mathcal{Z}^t)\allowbreak\textup{Pr}(A^t_2=a_2\!\mid\!\mathcal{Z}^t)... \allowbreak\textup{Pr}(A^t_N=a_N\!\mid\!\mathcal{Z}^t)$
is the probability that the $N$ sensors are in states $a_1,a_2,...,a_N$, respectively, at the end of slot $t$.
\end{defi}

Writing $\mathcal{Z}^t$ as $\mathcal{Z}^t= \{\mathcal{Z}^{t-1},n^t,A_{n^t}^{t-1} \}$, it is easy to show that the belief state $\bm{\Pi}^t$ is a sufficient statistic of $\mathcal{Z}^t$ \cite{SondikThesis}. In other words, $\bm{\Pi}^t$ summarizes all the information gained prior to the decision epoch at the end of slot $t$.

The belief state allows the POMDP to be formulated as a continuous-state MDP with states being the belief state $\bm{\Pi}^t$. The optimal policy in \eqref{eq:optpolicy} is then the solution to Bellman's dynamic programming recursion \cite{MDPBook}:
\begin{equation}\label{eq:Bellman}
\mu^*(\bm{\Pi}) \!=\! \arg\min_{n}\left\{\bar{A}_n(\bm{\Pi}) \!+\! \sum_{\bm{\Pi}^\prime} \textup{Pr}(\bm{\Pi}^\prime\!\mid\!\bm{\Pi},n) V^*(\bm{\Pi}^\prime)  \right\}.
\end{equation}
Note that in \eqref{eq:Bellman}, we have dropped the time index $t$ because the optimal policy is stationary, and
\begin{enumerate}
\item the relative cost-to-go function on $V^*(\bm{\Pi})$ is the difference between the total cost incurred by a system that starts with the state $\bm{\Pi}$ and the total cost incurred by a system that starts with a reference state over an infinite time horizon. If we set the equilibrium state as the reference state, $V^*(\bm{\Pi})$ is the extra cost incurred by the transient behavior of being in the state $\bm{\Pi}$.
\item $\bar{A}_n(\bm{\Pi})$ is the expected AoI incurred in one step by executing action $n$ (i.e., sample the $n$-th sensor), giving
\begin{eqnarray}\label{eq:expAoI}
\bar{A}_n(\bm{\Pi}) = \sum_{a_n=1}^{M} a_n \textup{Pr}(A_n=a_n\mid\mathcal{Z}^t).
\end{eqnarray}
\item $\textup{Pr}(\bm{\Pi}^\prime\mid\bm{\Pi},n)$ is the transition probability that the controller evolves from $\bm{\Pi}$ to $\bm{\Pi}^\prime$ if action $n$ is executed.
\end{enumerate}

In general, POMDPs are PSPACE hard problems to solve since the computation of the optimal policy \eqref{eq:Bellman} requires exponential computational complexity and memory. Therefore, we need to resort to suboptimal policies or algorithms for practical purposes.

\begin{rem}
As stated in the introduction, prior works have established a few sufficient conditions on the POMDP model under which the optimal policy is monotone in belief states (i.e., the optimal policy is a threshold policy) \cite{POMDPBook}. It is easy to verify that our problem does not satisfy the sufficient conditions. For example, one condition is that the transition probability matrix is ``totally positive of order 2'' (TP2), i.e., all the second-order minors of the transition matrix are non-negative. In our problem, however, $\bm{\mathcal{T}}_n$ is not TP2 since
$\textup{det}\begin{pmatrix}
q_n & p_n \\
q_n & 0
\end{pmatrix}<0$.
As a result, the monotonicity of the optimal policy (and hence inexpensive computation of the optimal policy) for our problem is unknown.
\end{rem}

\subsection{The greedy policy}
This paper explores the greedy policy to solve the POMDP. Compared with the optimal policy \eqref{eq:optpolicy} that minimizes the expected AoI over the infinite horizon, the greedy policy minimizes the expected AoI in the immediate step. That is, the greedy policy greedily samples the sensor with the minimum expected AoI in each slot. This policy is formally defined as follows:
\begin{defi}[the greedy policy $\mu^\prime$]
In each time slot, the greedy sampling policy $\mu^\prime$ instructs the AP to sample the sensor that yields the minimum expected AoI in one step, i.e.,
\begin{equation*}
\mu^\prime(\bm{\Pi}) \!=\! \arg\min_n \bar{A}_n(\bm{\Pi}) \!=\! \arg\min_n \sum_{a_n=1}^{M}\!\! a_n \textup{Pr}(A_n=a_n\!\mid\!\mathcal{Z}^t).
\end{equation*}
The performance of the greedy policy is
\begin{eqnarray}\label{eq:greedyperformance}
J(\mu^\prime)=\lim_{T\to\infty}\frac{1}{T}\sum_{t=0}^{T-1}\min_n \bar{A}_n(\bm{\Pi}^t),
\end{eqnarray}
i.e., the average sampled AoI over the infinite horizon.
\end{defi}

Despite its simple descriptions, the analysis of the greedy policy is non-trivial because the current decision will affect the performance going forward as expressed in \eqref{eq:greedyperformance}.
In other words, computing the greedy decision is simple; but computing the performance as a consequence of the decision is not trivial.  For performance-analytical purposes, we consider a relaxed greedy policy as an approximation to the greedy policy.  For the relaxed greedy policy, instead of restricting sampling to exactly one sensor for every time slot, we allow the AP to sample {\it on average} one sensor per time slot. That is, different numbers of sensors may be sampled in different time slots, but the average is one sensor per time slot. With this relaxation, we can decouple the original POMDP to $N$ independent POMDPs, each of which is associated with the sampling process of only one sensor.

\begin{defi}[the relaxed greedy policy $\hat{\mu}^\prime$]
In each time slot, the AP samples the $n$-th sensor if and only if its expected AoI is smaller than a constant $\eta$. Denote by $u_n^t$ an indication of whether the $n$-th sensor is sampled in time slot $t$: $u_n^t=1$ means ``sampled'' and $u_n^t=0$ means ``not sampled''. We have
\begin{eqnarray}
u_n^t = \mathbbm{1}_{\{\bar{A}^t_n<\eta\}}=
\begin{cases}
1& \text{if}~\bar{A}^t_n<\eta,\\
0& \text{if}~\bar{A}^t_n\geq\eta,
\end{cases}
\end{eqnarray}
where $\bar{A}_n^t$ is the expected AoI that can be obtained from the $n$-th sensor if the AP samples it (see equation \eqref{eq:expAoI}). The constant $\eta$ is chosen such that on average the AP samples one sensor per time slot. The performance of the relaxed greedy sampling policy $\hat{\mu}^\prime$ can then be expressed as
\begin{eqnarray}
\label{eq:relaxedgreedyperformance}
&&\hspace{-1cm} J(\hat{\mu}^\prime)=\lim_{T\to\infty}\frac{1}{T}\sum_{t=0}^{T-1}\sum_{n=1}^{N} \bar{A}^t_n u^t_n, \\
\label{eq:relaxedgreedycondition}
&&\hspace{-1cm} \textup{s. t.} \lim_{T\to\infty}\frac{1}{T}\sum_{t=0}^{T-1}\sum_{n=1}^{N} u^t_n = 1.
\end{eqnarray}
\end{defi}

With the greedy policy, the AP has to compare the expected AoI that can be sampled from each sensor and choose the sensor that yields the minimum expected AoI. With the relaxed greedy policy, on the other hand, the AP only needs to compare the expected AoI of each sensor with a constant $\eta$, and samples the sensors whose expected AoIs are smaller than $\eta$. By doing so, the sampling processes of the $N$ sensors are decoupled with each other, thereby making the relaxed greedy policy analyzable.
In the main body of this paper, we shall focus on the relaxed greedy policy and analyze its performance in terms of the expected sampled AoI over the infinite horizon.

\section{The decoupled POMDP}\label{sec:V}
The relaxed greedy policy is analyzable in that it allows the decoupling of the sampling process of the $N$ sensors. To understand the behavior of the relaxed greedy policy, it is important to study the sampling process of a single sensor.
To this end, this section considers a single-sensor sampling problem: the AP monitors the object via only one sensor. At the end of a slot $t$, the AP has to determine whether to ``sample'' or ``rest'' in the next slot. The state (AoI) of the sensor is determined by the MC in Fig.~\ref{fig:MC}, but the AP cannot directly observe the instantaneous state. Instead, the AP maintains a probability distribution $\bm{\pi}^t$ over the set of possible states of the sensor, and makes the sampling decisions (i.e., sample or rest) in consecutive slots based on $\bm{\pi}^t$.

It is evident that the single-sensor sampling problem itself constitutes a POMDP and the distribution $\bm{\pi}^t$ is the belief state of one sensor.

\begin{defi}[belief state of one sensor]
The belief state of one sensor is an $M$-dimensional posterior probability distribution $\bm{\pi}^t=\{\pi^t [k]:k=1,2,...,M\}$, where each entry $\pi^t[k]$ is the probability that the sensor is in state $k$ at the end of slot $t$, given the past decisions and observations $\{u^{t-\tau},a^{t-\tau-1}:\tau=0,1,2,3,...\}$, where $u^{t-\tau}$ and $a^{t-\tau-1}$ are the action of the AP and the observed AoI at the end of slot $t-\tau$, respectively.

The expected AoI that can be obtained from a belief state $\bm{\pi}$ is given by
\begin{equation*}
\bar{A}(\bm{\pi}) = \sum_{k=1}^{M}  k \bm{\pi}[m] \triangleq \bm{\pi Z_M},
\end{equation*}
where $\bm{Z_M} = [1,2,3,...,M]^\top$.
\end{defi}

\begin{figure}[t]
  \centering
  \includegraphics[width=0.8\columnwidth]{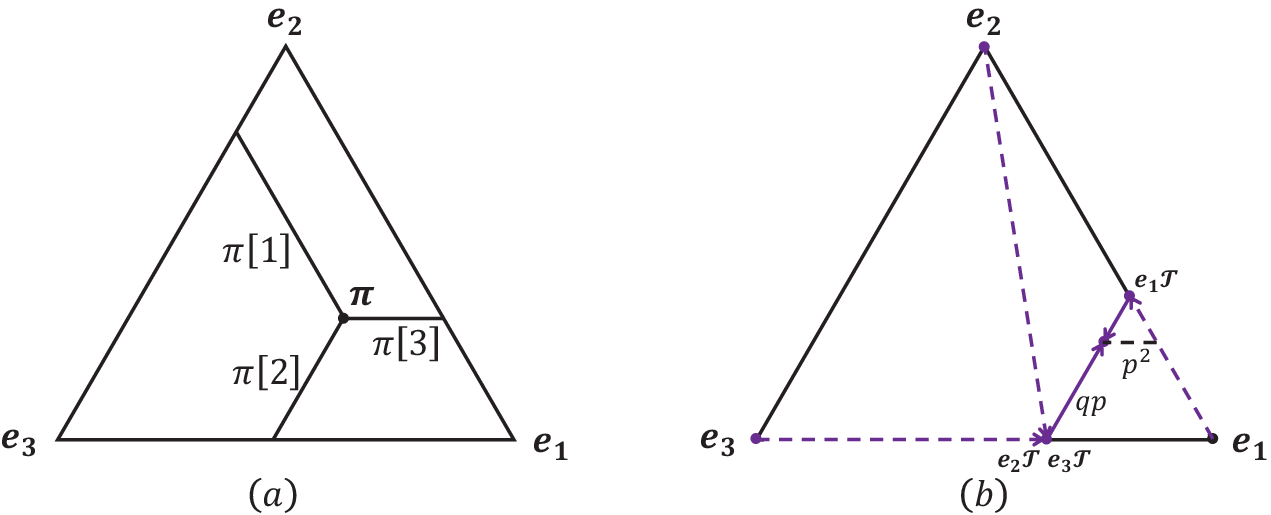}\\
  \caption{(a) The belief space for $M = 3$ is an equilateral triangle. When $M = 2$, the simplex is a unit line segment ($\bm{\pi}[1]+\bm{\pi}[2]=1$); when $M = 4$, the simplex is a tetrahedron. (b) The three evolution branches from $\bm{e}_1 \bm{\mathcal{T}}$, $\bm{e}_2 \bm{\mathcal{T}}$, and $\bm{e}_3 \bm{\mathcal{T}}$ to the steady state $\bm{h}=[q,qp,p^2]$ in two slots (one step).
  Specifically, $\bm{e}_1$, $\bm{e}_2$, and $\bm{e}_3$ are the unit belief states; $\bm{e}_1 \bm{\mathcal{T}}$, $\bm{e}_2 \bm{\mathcal{T}}$, and $\bm{e}_3 \bm{\mathcal{T}}$ are the initial states of the evolution branches. The purple dotted lines mean the transitions already happen when the AP observes, while the purple solid lines stand for the transitions on the evolution branches.
  }
\label{fig:simplex}
\end{figure}

The belief state $\bm{\pi}$ is a probability distribution. Thus, the space that $\bm{\pi}$ resides in is a unit simplex \cite{monotone3}. Denoted by $\bm{e}_m$ the unit belief state with one being in the $m$-th position. The unit belief states $\{\bm{e}_1,\bm{e}_2,...,\bm{e}_M \}$ are the vertices of the unit simplex. An example is shown in Fig.~\ref{fig:simplex}(a), where $M = 3$ and the belief space is an equilateral triangle.

At any belief state $\bm{\pi}^t$, the AP has two alternative actions: $u^t=1$ (sample) and $u^t=0$ (rest). For different actions, Lemma \ref{thm:lemma1} specifies the transitions of belief states in the belief space.

\begin{lem}[transitions of $\bm{\pi}$]\label{thm:lemma1}
The belief state $\bm{\pi}^t$ is Markovian. Given a belief state $\bm{\pi}^{t-1}$ and an action $u^t$, $\bm{\pi}^t$ is determined by
\begin{eqnarray}\label{eq:transitions}
\bm{\pi}^t = \begin{cases}
\bm{e}_k \bm{\mathcal{T}}  & \text{if}~u^t=1,\\
\bm{\pi}^{t-1} \bm{\mathcal{T}}  & \text{if}~u^t=0,
\end{cases}
\end{eqnarray}
where the transition matrix $\bm{\mathcal{T}}$ is given in \eqref{eq:matrix}, and we have assumed the AP observes an AoI $a^{t-1}=k\in\{1,2,3,...,M\}$ when $u^t=1$.
If the AP samples in slot $t$ ($u^t=1$) and then never sample, the belief state evolves from $\bm{e}_k \bm{\mathcal{T}}$ to a steady state $\bm{h}$ and stays in the steady state afterward. The steady-state distribution is given by $\bm{h}\bm{\mathcal{T}} = \bm{h}$.
\end{lem}

Lemma \ref{thm:lemma1} indicates that 1) the belief state will gradually evolve to a steady state if the AP does not sample the sensor; 2) once sampled, the belief state will be reset to one of the $M$ initial states $\{\bm{e}_k \bm{\mathcal{T}}:k=1,2,...,M\}$, depending on which state is sampled. In this light, we can divide the evolution of the belief states into $M$ evolution branches, each of which starts from an initial belief state $\bm{e}_k \bm{\mathcal{T}}$.

\begin{defi}[evolution branches]
Suppose the AP samples the sensor in slot $t+1$, observes an AoI $a^t=k$, and then never sample, we have
\begin{equation*}
\bm{\pi}^t = \bm{e}_k, ~~
\bm{\pi}^{t+i} = \bm{e}_k \bm{\mathcal{T}}^i.
\end{equation*}
Let $\bm{\pi}^{t+i}\triangleq \bm{\pi}_{k,i}$, we call the evolution $\{\bm{\pi}_{k,i}:i=1,2,3,...\}$ the $k$-th evolution branch of the belief states. Any belief state belongs to at least one evolution branch.
\end{defi}

Given the definition of the evolution branches, we can now characterize any belief state by two variates: the branch $k$ -- the observed AoI in the last sampling; and the elapsed slot $i$ -- how many slots elapsed since the last sampling.

\begin{lem}[belief states in an evolution branch]\label{thm:prop2}
The belief states in the $k$-th evolution branch are given by
\begin{equation}\label{eq:pi_ki}
\bm{\pi}_{k,i} =
\begin{bmatrix}
q,~qp,~ qp^2,~ ...,~ qp^{i-1},~ 0,~ ...,~ 0,~ p^i,~ 0,~ ...,~ 0
\end{bmatrix},
\end{equation}
for $i=1,2,3,...,M-1$, and the location of the last non-zero entry $p^i$ is $\min\{i+k,M\}$. When $i\geq M-1$, the evolution branch enters the steady state and the belief state no longer changes, giving $
\bm{\pi}_{k,i} = \bm{h} =
[q,\allowbreak qp,\allowbreak qp^2,\allowbreak \cdots,\allowbreak qp^{M-2},\allowbreak p^{M-1}]$.
\end{lem}

Lemma \ref{thm:prop2} can be proven by induction. First, it is easy to verify that
$\bm{\pi}_{k,1}=\bm{e}_k \bm{\mathcal{T}}=[q,\allowbreak 0,\allowbreak ...,0,\allowbreak p,\allowbreak 0,\allowbreak ...,\allowbreak 0]$,
and the location of $p$ is $\min\{k+1,M\}$. This satisfies \eqref{eq:pi_ki}.

Let $i=1,2,3,...,M-2$. If $\bm{\pi}_{k,i}$ satisfies \eqref{eq:pi_ki}, i.e.,
$\bm{\pi}_{k,i}=[q,\allowbreak qp,\allowbreak qp^2,...,~qp^{i-1},\allowbreak 0,...,0,\allowbreak p^i,\allowbreak 0,\allowbreak ...,\allowbreak 0]$,
with $p^i$ being in the $\min\{i+k,M\}$-th position, then we have
$\bm{\pi}_{k,i+1} \allowbreak=\bm{\pi}_{k,i} \bm{\mathcal{T}} =\allowbreak [q,qp,qp^2,\allowbreak ...,\allowbreak qp^{i-1},qp^{i},\allowbreak 0,...,0,\allowbreak p^{i+1},\allowbreak 0,...,0]$,
with $p^{i+1}$ being in the $\min\{i+k+1,M\}$-th position. This satisfies \eqref{eq:pi_ki}.

Let $i= M-1$, we have $\bm{\pi}_{k,M-1}= [q,\allowbreak qp,\allowbreak qp^2,...,\allowbreak qp^{M-2},\allowbreak p^{M-1}]$ and
$\bm{\pi}_{k,M-1} = \bm{\pi}_{k,M-1} \bm{\mathcal{T}}$.
That is, $\bm{\pi}_{k,M-1}$ satisfies $\bm{h}=\bm{h}\bm{\mathcal{T}}$ and hence is the steady state. The belief state no longer changes for $i\geq M-1$.

An important result of Lemma \ref{thm:prop2} is that it takes a finite number of $M - 1$ slots for an initial state $\bm{\pi}_{k,1}$ to evolve to the steady state $\bm{h}$ if the AP does not sample the sensor. A simple example is shown in Fig.~\ref{fig:simplex}(b), wherein $M = 3$. The three evolution paths start from $\bm{e}_1 \bm{\mathcal{T}}$, $\bm{e}_2 \bm{\mathcal{T}}$, and $\bm{e}_3 \bm{\mathcal{T}}$, respectively, and evolve to the steady state $\bm{h}=[q,qp,p^2 ]$ in $M - 1 = 2$ slots (i.e., takes only one step). If the AP samples the sensor during the evolution, the belief state is reset to one of the $\{\bm{e}_1 \bm{\mathcal{T}}$, $\bm{e}_2 \bm{\mathcal{T}}$, $\bm{e}_3 \bm{\mathcal{T}}\}$, as per \eqref{eq:transitions}.

For the greedy and the relaxed greedy policies, a critical statistic of a belief state is the expected AoI $\bar{A}(\bm{\pi})$ that can be obtained if the AP samples the sensor in state $\bm{\pi}$. With the relaxed greedy policy, for example, the AP compares the expected AoI of a state with a constant $\eta$: if $\bar{A}(\bm{\pi})<\eta$, the AP samples the sensor; and if $\bar{A}(\bm{\pi})\geq\eta$, the AP does not sample the sensor. In other words, the expected AoI measures the quality of a belief state.

Given the useful representation of the belief state in Lemma \ref{thm:prop2}, the expected AoI of a belief state is simply a function of $k$ and $i$. Next, we analyze how the expected AoI evolves in each evolution branch. Proposition~\ref{thm:evolutions} below summarizes our main results in the section.

\begin{prop}[evolutions of the expected AoI]\label{thm:evolutions}
Denote by $\bar{A}_{k,i}$ the expected AoI that can be obtained in the belief state $\bm{\pi}_{k,i}$. We have,
\begin{eqnarray}\label{eq:evolutions}
\bar{A}_{k,i} = \bm{\pi}_{k,i} \bm{Z_M} = \frac{1-p^i}{1-p} - p^i i + p^i \min\{i+k,M\},
\end{eqnarray}
where $k=1,2,...,M$ and $i=1,2,3,...,M-1$. In particular, the expected AoI of the steady state
\begin{eqnarray}\label{eq:steadystateAoI}
\bar{\bm{h}}=\bar{A}_{k,M-1} = \frac{1-p^M}{1-p}.
\end{eqnarray}
For different evolution branches $k=1,2,...,M$, the evolution from $\bar{A}_{k,1}$ to $\bar{A}_{k,M-1}$ goes through two phases.
\begin{enumerate}[\hspace{1pt}1)]
\item Phase 1, $1\leq i\leq M\!-\! k$: if $k\leq \frac{1}{1-p}$, $\bar{A}_{k,1}$ increases monotonically to $\bar{A}_{k,M-k}$; if $k>\frac{1}{1-p}$, $\bar{A}_{k,1}$ decreases monotonically to $\bar{A}_{k,M-k}$.
\item Phase 2, $M\!-\! k<i\leq M\!-\! 1$: $\bar{A}_{k,M-k}$ decreases monotonically to$\bar{A}_{k,M-1}$. In particular, $\bar{A}_{k,i}$ is irrelevant to $k$ in this process, i.e., $\bar{A}_{1,i}=\bar{A}_{2,i}=...=\bar{A}_{M,i}$.
\end{enumerate}
Whenever the AP samples the sensor, the expected AoI is reset to an initial expected AoI $\bar{A}_{k^\prime,1}$ if the sampled AoI is $k^\prime\allowbreak=1,\allowbreak 2,\allowbreak ...,M$.
\end{prop}

\begin{NewProof}
Eq. \eqref{eq:evolutions} and \eqref{eq:steadystateAoI} follows from \eqref{eq:transitions} and \eqref{eq:pi_ki}. In particular,

1) If $i+k\leq M$, we have $\bar{A}_{k,i}=\frac{1-p^i}{1-p}-p^i i+p^i (i+k)=\frac{1-p^i}{1-p}+kp^i$.
Differentiate $\bar{A}_{k,i}$ with respect to $i$ gives us
\begin{eqnarray}
\frac{\partial \bar{A}_{k,i}}{\partial i} = p^i \left( k-\frac{1}{1-p} \right) \ln p,
\end{eqnarray}
where $p^i\ln p<0$ for $p\in (0,1)$.
Thus, if $k>\frac{1}{1-p}$, $\bar{A}_{k,i}$ is a decreasing function of $i$, and $k\leq\frac{1}{1-p}$, $\bar{A}_{k,i}$ is a (weakly) increasing function of $i$.

2) If $i+k>M$, we have
$\bar{A}_{k,i}=\frac{1-p^i}{1-p}-p^i i+p^i M$.
Since $\bar{A}_{k,i}$ is not a function of $k$, we have $\bar{A}_{1,i}=\bar{A}_{2,i}=...=\bar{A}_{M,i}$.
Differentiate $\bar{A}_{k,i}$ with respect to $i$ gives us
\begin{eqnarray}
\frac{\partial \bar{A}_{k,i}}{\partial i} = -p^i \left(  \frac{1}{1-p} + \frac{1}{\ln p} + i - M \right) \ln p,
\end{eqnarray}
where $-p^i \ln p > 0$, $\frac{1}{1-p} + \frac{1}{\ln p}<0$ for $p\in (0,1)$, and $i-M<0$. Thus, $\bar{A}_{k,i}$ is a decreasing function of $i$.

Overall, 1) For $k\leq \frac{1}{1-p}$, $\bar{A}_{k,1}$  first increases monotonically to $\bar{A}_{k,M-k}$ , and then decreases monotonically to $\bar{A}_{k,M-1}$ . 2) For $k>\frac{1}{1-p}$, $\bar{A}_{k,1}$ decreases monotonically to $\bar{A}_{k,M-1}$.
\end{NewProof}

\begin{cor}
If $k_1\geq k_2$, $\bar{A}_{k_1,i}\geq \bar{A}_{k_2,i}$.
\end{cor}

\begin{figure}[t]
  \centering
  \includegraphics[width=0.8\columnwidth]{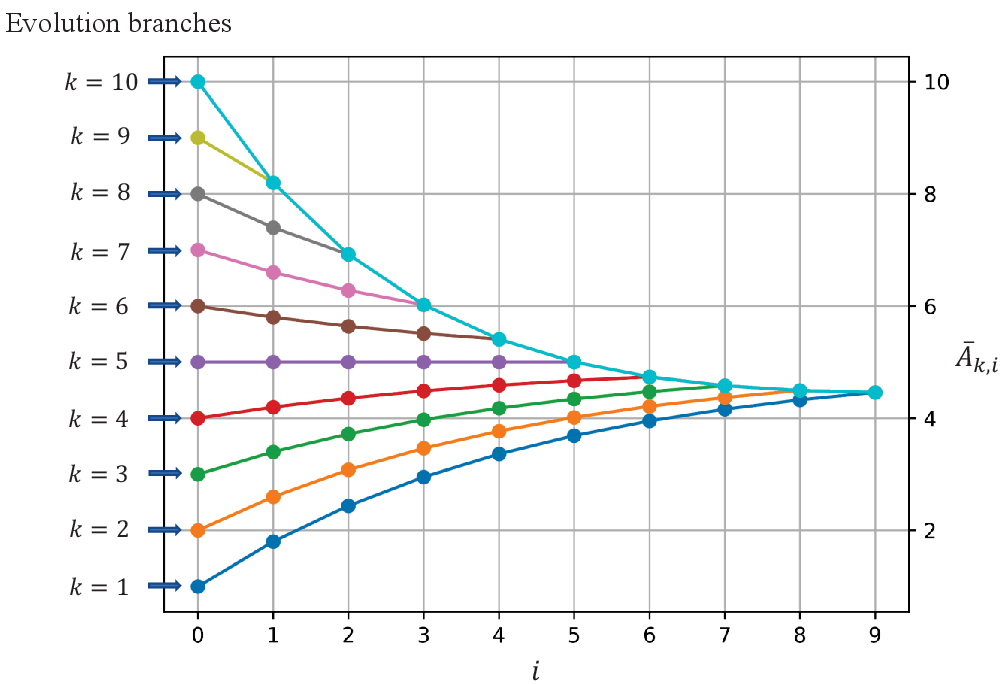}\\
  \caption{An example of the evolutions of the expected AoI for each evolution branch, wherein $M = 10$, $p = 0.8$.}
\label{fig:example}
\end{figure}

Proposition \ref{thm:evolutions} is a cornerstone of our analysis in Section \ref{sec:VI}. We next give an example in Fig.~\ref{fig:example} to show visually the evolutions of the expected AoI in each evolution branch. In Fig.~\ref{fig:example}, we set $M = 10$, $p = 0.8$, and plot $\bar{A}_{k,i}$ as a function of $i$ for different $k$ (i.e., one curve in Fig.~\ref{fig:example} corresponds to one evolution branch). As can be seen, 1) from $\bar{A}_{k,1}$, it takes $M-1=9$ slots for the expected AoI to evolve to $\bar{\bm{h}}=\bar{A}_{k,M-1}$; 2) for $k\leq\frac{1}{1-p} = 5$, $\bar{A}_{k,1}$ increases monotonically to $\bar{A}_{k,M-k}$ and then decreases monotonically to $\bar{A}_{k,M-1}$; for $k>5$, $\bar{A}_{k,1}$ decreases monotonically to $\bar{A}_{k,M-1}$. These observations are consistent with Proposition \ref{thm:evolutions}.

\section{The Relaxed Greedy Policy}\label{sec:VI}
Section \ref{sec:V} dissects the decoupled POMDP associated with the sampling process of one sensor and studies the evolutions of the expected AoI in different evolution branches. With the results established in Section \ref{sec:V}, we are ready to analyze the performance of the relaxed greedy sampling.

\subsection{Thresholds for the Evolution Branches}
Assuming that the AP never samples the sensor, Proposition \ref{thm:evolutions} describes the evolution of the expected AoI in each evolution branch -- in a finite number of $M - 1$ slots, the expected AoI evolves from one of the $M$ initial expected AoI $\{\bar{A}_{k,1}:k=1,2,...,M\}$ to a steady-state expected AoI $\bar{A}_{k,M-1}$ following \eqref{eq:evolutions}. If the AP samples the sensor in this process, however, the expected AoI is reset to one of the $M$ initial expected AoI, and the cycle continues.

With the relaxed greedy policy, the AP samples the sensor if $\bar{A}_{k,i}<\eta$. Thus, the $k$-th evolution branch terminates at $i=\gamma_k$, where
\begin{eqnarray}\label{eq:thres}
\gamma_k=\inf_\gamma\{\gamma: \bar{A}_{k,\gamma_k}<\eta\}.
\end{eqnarray}
Visually, one can think of $\bar{A}_{k,i}=\eta$ as a straight line in Fig. \ref{fig:example}. For the $k$-th branch, $\gamma_k$ is the $x$-coordinate (i.e., the number of elapsed slots $i$) of the first point whose $y$-coordinate (i.e., the expected AoI) is smaller than $\eta$. At this point, the AP will sample the sensor and the expected AoI will evolve back to one of the $M$ initial points.

For a single sensor, the threshold $\gamma_k$ for each evolution branch is characterized in Theorem \ref{thm:thresholds}.

\begin{thm}[thresholds for the evolution branches]\label{thm:thresholds}
Consider a single sensor. With the relaxed greedy policy, the AP samples the sensor if and only if the expected AoI obtained from this sensor is less than a constant $\eta$.
\begin{enumerate}[\hspace{1pt}1)]
\item If $\eta$ is larger than the initial expected AoI of the $M$-th evolution branch, i.e. $\eta>\bar{A}_{M,1}= q + M p$, we have $\gamma_k=1,~\forall~k$.
\item If $\eta$ is smaller than or equal to the initial value of the $M$-th branch but larger than the steady-state expected AoI, i.e., $\bar{\bm{h}}<\eta\leq\bar{A}_{M,1}$, we have
\begin{equation*}
\left\{
\begin{array}{lll}
\gamma_k=1, \hspace{0.5cm} &&\hspace{-5.7cm}\text{for}~\left\{k:\bar{A}_{k,1}=q+\min\{k+1,M\}p<\eta \right\}\\
\gamma_k=\left\lceil\frac{1}{\ln p}W_0\left(\psi(\eta) p^{\psi(M)}\ln p\right) - \psi(M)\right\rceil, \\
&&\hspace{-6cm}\text{for}~\left\{k:\bar{A}_{k,1}\geq\eta,k\leq\frac{1}{1-p}\right\} \\
&&\hspace{-6cm}\text{or}~\left\{k:\bar{A}_{k,1}\geq\eta,k>\frac{1}{1-p},\bar{A}_{k,M-k-1}>\eta \right\},\\
\gamma_k=\left\lceil \log_p\left(\frac{1-\eta(1-p)}{1-k(1-p)}\right) \right\rceil, \\
&&\hspace{-6cm}\text{for}~\left\{k:\bar{A}_{k,1}\geq\eta,k>\frac{1}{1-p},\bar{A}_{k,M-k-1}\leq\eta \right\}
\end{array}
\right.
\end{equation*}
where $W_0(\star)$ is the principal branch of a Lambert W function and $\psi(x)=\frac{1}{1-p}-x$.
\item If $\eta$ is smaller than or equal to the steady-state expected AoI, i.e., $0\leq\eta\leq\bar{\bm{h}}$, we have
\begin{eqnarray*}
\gamma_k =\begin{cases}
\infty& \text{if}~\eta\leq\bar{A}_{k,1},\\
\left\lceil \log_p\left(\frac{1-\eta(1-p)}{1-k(1-p)}\right) \right\rceil  & \text{if}~\eta>\bar{A}_{k,1}.
\end{cases}
\end{eqnarray*}
In this case, the AP would never sample the sensor if at some point in time the AP samples an AoI $k^\prime\in\{k^\prime:\eta\leq\bar{A}_{k^\prime,1}\}$ because $\gamma_{k^\prime}=\infty$.
\end{enumerate}
\end{thm}

\begin{NewProof}
See Appendix \ref{sec:AppA}.
\end{NewProof}

\begin{cor}
For a single sensor, $1\leq \gamma_1 \leq \gamma_2\leq ...\leq \gamma_M$.
\end{cor}

\subsection{Performance of the Relaxed Greedy Policy}
With the relaxed greedy policy, the sampling processes of the $N$ sensors are decoupled because the AP only needs to compare the state of each sensor with a constant $\eta$ to determine whether to sample it or not. Following \eqref{eq:relaxedgreedyperformance} and \eqref{eq:relaxedgreedycondition}, the performance of the relaxed greedy policy, i.e., the average sampled AoI over the infinite horizon, can be rewritten as
\begin{eqnarray}
\label{eq:relaxedgreedyperformance2}
&&\hspace{-1cm} J(\hat{\mu}^\prime)=\lim_{T\to\infty}\frac{1}{T}\sum_{t=0}^{T-1}\sum_{n=1}^{N} \bar{A}^t_n \mathbbm{1}_{\{\bar{A}^t_n<\eta\}}=\sum_{n=1}^{N} \bar{R}_n, \\
\label{eq:relaxedgreedycondition2}
&&\hspace{-1cm} \textup{s. t.} \lim_{T\to\infty}\frac{1}{T}\sum_{t=0}^{T-1}\sum_{n=1}^{N} \mathbbm{1}_{\{\bar{A}^t_n<\eta\}} = \sum_{n=1}^{N} \bar{d}_n = 1.
\end{eqnarray}
where we have defined $\bar{R}_n$ as the average AoI that can be obtained from the $n$-th sensor per slot, and $\bar{d}_n$ as the average number of times that the $n$-th sensor is sampled per slot.

These two variables can be further manipulated as
\begin{eqnarray}
\label{eq:Rbar}
&&\hspace{-1cm} \bar{R}_n=\lim_{T\to\infty}\frac{1}{T}\sum_{t=0}^{T-1} \bar{A}^t_n \mathbbm{1}_{\{\bar{A}^t_n<\eta\}}\triangleq \lim_{T\to\infty} \frac{R_n(T)}{T}, \\
\label{eq:dbar}
&&\hspace{-1cm} \bar{d}_n = \lim_{T\to\infty}\frac{1}{T}\sum_{t=0}^{T-1} \mathbbm{1}_{\{\bar{A}^t_n<\eta\}} \triangleq \lim_{T\to\infty} \frac{d_n(T)}{T},
\end{eqnarray}
where $R_n(T)$ is defined as the sum of the expected AoI obtained from the $n$-th sensor in $T$ slots, and $d_n(T)$ is the average number of times that the $n$-th sensor is sampled in $T$ slots.

The train of thought to derive the performance of the relaxed greedy policy is as follows.
\begin{description}
\item[Step 1:] $~$ Derive the $d_n(T)$ and $\bar{d}_n$ in \eqref{eq:dbar} for each sensor;
\item[Step 2:] $~$ Given the constraint \eqref{eq:relaxedgreedycondition2}, find the highest $\eta^*$ such that the AP samples one sensor per slot on average;
\item[Step 3:] $~$ Given $\eta^*$, derive the $R_n(T)$ and $\bar{R}_n$ in \eqref{eq:Rbar}, and compute $J(\hat{\mu}^\prime)$ following \eqref{eq:relaxedgreedyperformance2} as a sum of $\bar{R}_n$.
\end{description}

\subsubsection{Deriving the $d_n(T)$ and $\bar{d}_n$ for each sensor given a constant $\eta$}
Given a constant $\eta$, we first analyze $d_n(T)$ and $\bar{d}_n$  for a single sensor. To simplify the notations, we drop the subscript $n$ in this subsection.

Consider a sampling trajectory of one sensor. Suppose the AoI of the sensor is initialized to $a^t=k$ at the end of slot $t=0$, and the belief state of the sensor is initialized to $\bm{\pi}^0=\bm{e}_k$ at the AP. For a given $\eta$, Theorem \ref{thm:thresholds} indicates that the $k$-th evolution branch of the belief state will last for $\gamma_k$ slots before the AP samples the sensor. Denote by $d(k,T)$ the average number of times that the sensor is sampled in $T$ slots (i.e., we add index $k$ onto $d(T)$ to denote that the sampling trajectory starts from $a^0=k$), we have
\begin{equation*}
d(k,T)=0,~T\in[0,\gamma_k-1]; ~~~~ d(k,\gamma_k)=1.
\end{equation*}

In slot $\gamma_k$, the AP samples the sensor, and the belief state evolves back to one of the $M$ initial belief states $\bm{\pi}_{j,1}$, $j=1,2,...,M$ with probability $\bm{\pi}_{k,\gamma_k}[j]$. As a result, in slot $T=\gamma_k+\tau$, $d(k,\gamma_k+\tau)$ is defined by the following recurrence relation
\begin{eqnarray}\label{eq:recurrence}
d(k,\gamma_k+\tau) = 1 \!+\! \bm{\pi}_{k,\gamma_k} \left[d(1,\tau),d(2,\tau),...,d(M,\tau)\right]^\top,
\end{eqnarray}
where $\tau\in[1,\infty)$; $\{d(j,\tau):j=1,2,...,M\}$ on the RHS is the average number of times that the sensor is sampled in $\tau$ slots if the trajectory starts from $a^0=j$; and $(*)^\top$ denotes the transpose of a vector.

For different $k=1,2,...,M$, \eqref{eq:recurrence} defines a set of $M$ recurrence relations, the matrix form of which can be written as
\begin{eqnarray}\label{eq:recurrenceMatrix}
\begin{bmatrix}
\begin{smallmatrix}
d(1,\gamma_1+\tau)\\
d(2,\gamma_2+\tau)\\
\cdots\\
d(M,\gamma_M+\tau)\\
\end{smallmatrix}
\end{bmatrix}=
\begin{bmatrix}
\begin{smallmatrix}
1\\
1\\
\cdots\\
1\\
\end{smallmatrix}
\end{bmatrix} +
\begin{bmatrix}
\begin{smallmatrix}
\cdots & \bm{\pi}_{1,\gamma_1} & \cdots\\
\cdots & \bm{\pi}_{2,\gamma_2} & \cdots\\
\cdots & \cdots & \cdots\\
\cdots & \bm{\pi}_{M,\gamma_M} & \cdots
\end{smallmatrix}
\end{bmatrix}\!\!
\begin{bmatrix}
\begin{smallmatrix}
d(1,\tau)\\
d(2,\tau)\\
\cdots\\
d(M,\tau)\\
\end{smallmatrix}
\end{bmatrix}.
\end{eqnarray}
These recurrence relations are very important in that they define $d(k,T)$ recursively -- provided that $\{d(j,\tau):j=1,2,...,M\}$ are known, $\{d(k,\gamma_k+\tau):k=1,2,...,M\}$ can be computed accordingly.

\begin{prop}[]\label{thm:propd}
For a given $\eta$ and the corresponding thresholds $\{\gamma_k:k=1,2,...,M\}$ associated with the $M$ evolution branches, the average number of times that a sensor is sampled per slot
\begin{eqnarray}\label{eq:bard}
\bar{d} = \frac{\text{det}(\bm{B_M})}{\text{det}(\bm{B})},
\end{eqnarray}
where the matrices $\bm{B_M}$ and $\bm{B}$ are given by
\begin{eqnarray*}
&&\bm{B_M} =
\begin{bmatrix}
\begin{smallmatrix}
\bm{\pi}_{1,\gamma_1}[1]-1  &  \bm{\pi}_{1,\gamma_1}[2] & ... & \bm{\pi}_{1,\gamma_1}[M-1] & 1\\
\bm{\pi}_{2,\gamma_2}[1]  &  \bm{\pi}_{2,\gamma_2}[2]-1 & ... & \bm{\pi}_{2,\gamma_2}[M-1] & 1\\
...                         &  ...                      & ... & ...                       & ...\\
\bm{\pi}_{M,\gamma_M}[1]-1  &  \bm{\pi}_{M,\gamma_M}[2] & ... & \bm{\pi}_{M,\gamma_M}[M-1] & 1
\end{smallmatrix}
\end{bmatrix}, \\
&&\bm{B} =
\begin{bmatrix}
\begin{smallmatrix}
\bm{\pi}_{1,\gamma_1}[1]-1  &  \bm{\pi}_{1,\gamma_1}[2] & ... & \bm{\pi}_{1,\gamma_1}[M-1] & \gamma_1\\
\bm{\pi}_{2,\gamma_2}[1]  &  \bm{\pi}_{2,\gamma_2}[2]-1 & ... & \bm{\pi}_{2,\gamma_2}[M-1] & \gamma_2\\
...                         &  ...                      & ... & ...                       & ...\\
\bm{\pi}_{M,\gamma_M}[1]-1  &  \bm{\pi}_{M,\gamma_M}[2] & ... & \bm{\pi}_{M,\gamma_M}[M-1] & \gamma_M
\end{smallmatrix}
\end{bmatrix}.
\end{eqnarray*}
\end{prop}

\begin{NewProof}
For different starting state $k=1,2,...,M$, $\bar{d}=d(k,\tau)/\tau$ converges to the same value because the average number of times that a sensor is sampled per slot is irrelevant to the initial state that the sampling trajectory starts with.

Let $\lim_{T\to\infty}\frac{\partial d(k,T)}{\partial T}=\alpha$, we have
\begin{eqnarray}\label{eq:AppD1}
d(k,T) = \alpha T + b(k),
\end{eqnarray}
where $\{b(k)\}$ are constants for different $k=1,2,...,M$. Notice that our target value $\bar{d} \allowbreak= \lim_{T\to\infty}\allowbreak\frac{d(k,T)}{T} \allowbreak= \alpha$. Our aim is then to derive $\alpha$.

Substituting \eqref{eq:AppD1} into \eqref{eq:recurrenceMatrix} gives us
\begin{equation*}
\begin{bmatrix}
\begin{smallmatrix}
\alpha(\gamma_1+\tau)+b(1)\\
\alpha(\gamma_2+\tau)+b(2)\\
\cdots\\
\alpha(\gamma_M+\tau)+b(M)\\
\end{smallmatrix}
\end{bmatrix}\!\! =\!\!
\begin{bmatrix}
\begin{smallmatrix}
1\\
1\\
\cdots\\
1\\
\end{smallmatrix}
\end{bmatrix} \!\!+\!\!
\begin{bmatrix}
\begin{smallmatrix}
\cdots & \bm{\pi}_{1,\gamma_1} & \cdots\\
\cdots & \bm{\pi}_{2,\gamma_2} & \cdots\\
\cdots & \cdots & \cdots\\
\cdots & \bm{\pi}_{M,\gamma_M} & \cdots
\end{smallmatrix}
\end{bmatrix}\!\!
\begin{bmatrix}
\begin{smallmatrix}
\alpha \tau+b(1)\\
\alpha \tau+b(2)\\
\cdots\\
\alpha \tau+b(M)\\
\end{smallmatrix}
\end{bmatrix}.
\end{equation*}
After some manipulations, we have
\begin{eqnarray}\label{eq:AppD4}
\left(\begin{bmatrix}
\begin{smallmatrix}
\cdots & \bm{\pi}_{1,\gamma_1} & \cdots\\
\cdots & \bm{\pi}_{2,\gamma_2} & \cdots\\
\cdots & \cdots & \cdots\\
\cdots & \bm{\pi}_{M,\gamma_M} & \cdots
\end{smallmatrix}
\end{bmatrix} -
\textup{eye}(M)\right)
\begin{bmatrix}
\begin{smallmatrix}
b(1)\\
b(2)\\
\cdots\\
b(M)\\
\end{smallmatrix}
\end{bmatrix} =
\begin{bmatrix}
\begin{smallmatrix}
\alpha\gamma_1 - 1\\
\alpha\gamma_2 - 1\\
\cdots\\
\alpha\gamma_M - 1\\
\end{smallmatrix}
\end{bmatrix},
\end{eqnarray}
where $\textup{eye}(M)$ denotes a unit matrix of size $M\times M$. We can then derive $\alpha$ by finding a particular solution of \eqref{eq:AppD4}. Let $b(M)=0$, we have
\begin{equation*}
\begin{bmatrix}
\begin{smallmatrix}
\bm{\pi}_{1,\gamma_1}[1]-1  &  \bm{\pi}_{1,\gamma_1}[2] & ... & \bm{\pi}_{1,\gamma_1}[M-1] & \gamma_1\\
\bm{\pi}_{2,\gamma_2}[1]  &  \bm{\pi}_{2,\gamma_2}[2]-1 & ... & \bm{\pi}_{2,\gamma_2}[M-1] & \gamma_2\\
...                         &  ...                      & ... & ...                       & ...\\
\bm{\pi}_{M,\gamma_M}[1]-1  &  \bm{\pi}_{M,\gamma_M}[2] & ... & \bm{\pi}_{M,\gamma_M}[M-1] & \gamma_M
\end{smallmatrix}
\end{bmatrix}\!\!\!
\begin{bmatrix}
\begin{smallmatrix}
b(1)\\
b(2)\\
\cdots\\
b(M-1)\\
-\alpha
\end{smallmatrix}
\end{bmatrix}\!\!\!=\!\!
\begin{bmatrix}
\begin{smallmatrix}
-1 \\
-1 \\
... \\
-1 \\
\end{smallmatrix}
\end{bmatrix}.
\end{equation*}
The $\bar{d}$ in \eqref{eq:bard} then follows from the Cramer's rule.
\end{NewProof}

\subsubsection{Finding the optimal $\eta^*$}
Given any $\eta$, we can compute the $\bar{d}$ for each sensor following \eqref{eq:bard}. As specified in Theorem \ref{thm:thresholds}, if the steady-state expected AoI of a sensor is larger than or equal to $\eta$ (i.e., $0\leq\eta\leq\bar{\bm{h}}$), the AP will never sample this sensor at some point in time (i.e., $\gamma_k=\infty$ and $\bar{d}=0$). This means, in terms of satisfying the constraint \eqref{eq:relaxedgreedycondition2}, we only need to consider the sensors whose steady-state expected AoI is smaller than the $\eta$, because the $\bar{d}$ of other sensors are zero (they will never be sampled by the AP).

In light of this, we sort the $N$ sensors so that sensors with larger indexes have larger steady-state expected AoI.\footnote{Eq. \eqref{eq:steadystateAoI} indicates that larger $p$ gives the sensor larger $\hat{\bm{h}}$. Thus, the $N$ sensors are actually sorted by the value of $p$ (sensors with larger $p$ have larger indexes).} Given a constant $\eta$, we compute the $\bar{d}$ for each sensor following \eqref{eq:bard}, and denote them by $\{\bar{d}_1(\eta),\allowbreak\bar{d}_2(\eta),\allowbreak...,\bar{d}_N(\eta)\}$. The set of sensors that can be sampled by the AP is $\{n:\bar{\bm{h}}_n<\eta\}$.

The optimal $\eta^*$ is then found by
\begin{eqnarray}
\label{eq:dhat}
&&\hspace{-0.5cm}\hat{d}(\eta) = \sum_{n=1}^{\sup\{n:\bar{\bm{h}}_n<\eta\}} \bar{d}_n(\eta), \\
\label{eq:opteta}
&&\hspace{-0.5cm} \eta^* =\arg\min_{\eta} \left|\hat{d}(\eta) - 1 \right|.
\end{eqnarray}
That is, given an $\eta$, we first calculate $\hat{d}(\eta)$ following \eqref{eq:dhat}, i.e., the sum of $\bar{d}_n(\eta)$ for sensors whose steady-state expected AoI is smaller than the $\eta$. As per \eqref{eq:relaxedgreedycondition2}, we have to find the $\eta^*$ that yields $\hat{d}(\eta^*)=1$. A caveat here is that such $\eta^*$ may not exist. Thus, we choose the $\eta^*$ to minimize the difference between $\hat{d}(\eta)$ and $1$.\footnote{It is worth noting that $\hat{d}_n(\eta)$ is in general a discontinuous function of $\eta$. When we gradually increase $\eta$ to minimize the difference between $\hat{d}_n(\eta)$ and $1$, the found $\eta^*$ often leads to a $\hat{d}_n(\eta^*)$ that is smaller than $1$. As a result, the performance of the relaxed greedy policy is often slightly better than the greedy policy as slightly fewer sensors are sampled.} This gives us \eqref{eq:opteta}.

\subsubsection{Compute $R_n(T)$, $\bar{R}_n$, and $J(\hat{\mu}^\prime)$ for the relaxed greedy policy}
In steps 1 and 2, we have determined the optimal $\eta^*$ and the set of sensors that can be sampled by the AP, i.e., $\{n:\bar{\bm{h}}_n<\eta^*\}$. For each sensor in this set, $R(T)$ (i.e., the sum of the expected AoI obtained from a sensor in $T$ slots, the subscript $n$ is omitted) can be computed by the following recurrence relation.
Considering a sampling trajectory of one sensor, for $\tau\in[1,\infty)$,
\begin{eqnarray}\label{eq:recurrenceMatrix2}
\begin{bmatrix}
\begin{smallmatrix}
R(1,\gamma^*_1+\tau)\\
R(2,\gamma^*_2+\tau)\\
\cdots\\
R(M,\gamma^*_M+\tau)\\
\end{smallmatrix}
\end{bmatrix}\!\! =\!\!
\begin{bmatrix}
\begin{smallmatrix}
\bar{A}_{1,\gamma^*_1}\\
\bar{A}_{2,\gamma^*_2}\\
\cdots\\
\bar{A}_{M,\gamma^*_M}\\
\end{smallmatrix}
\end{bmatrix} \!\!+\!\!
\begin{bmatrix}
\begin{smallmatrix}
\cdots & \bm{\pi}_{1,\gamma^*_1} & \cdots\\
\cdots & \bm{\pi}_{2,\gamma^*_2} & \cdots\\
\cdots & \cdots & \cdots\\
\cdots & \bm{\pi}_{M,\gamma^*_M} & \cdots
\end{smallmatrix}
\end{bmatrix}\!\!
\begin{bmatrix}
\begin{smallmatrix}
R(1,\tau)\\
R(2,\tau)\\
\cdots\\
R(M,\tau)\\
\end{smallmatrix}
\end{bmatrix},
\end{eqnarray}
where we have added index $k$ onto $R(T)$ to denote that the sampling trajectory starts from $a^0=k$. This relation is derived in a similar fashion to \eqref{eq:recurrenceMatrix}.
Suppose the AoI of the sensor is initialized to $a^t=k$ at the end of slot $t=0$, and the belief state of the sensor is initialized to $\bm{\pi}^0=\bm{e}_k$ at the AP.
For the optimal $\eta^*$, we can compute the optimal threshold $\gamma_k^*$ for the $k$-th evolution branch. The AP will sample the sensor at $T=\gamma_k^*$, thus we have
\begin{eqnarray*}
R(k,T) = 0,~T\in[0,\gamma_k^* - 1],~~~R(k,\gamma_k^*) = \bar{A}_{k,\gamma^*_k},
\end{eqnarray*}
where $\bar{A}_{k,\gamma_k^*}$ is the expected AoI obtained at slot $T=\gamma_k^*$. The belief state then evolves back to one of the $M$ initial belief states $\bm{\pi}_{j,1}$, $j=1,2,...,M$ with probability $\bm{\pi}_{k,\gamma_k^*}[j]$. Eq. \eqref{eq:recurrenceMatrix2} thus follows.

\begin{prop}[]\label{thm:propR}
For the optimal $\eta$ and the corresponding thresholds $\{\gamma^*_k:k=1,2,...,M\}$ associated with the $M$ evolution branches, the average AoI that can be obtained from a sensor per slot
\begin{eqnarray*}
\bar{R} = \frac{\text{det}(\bm{B^*_M})}{\text{det}(\bm{B^*})},
\end{eqnarray*}
where the matrices $\bm{B^*_M}$ and $\bm{B^*}$ are given by
\begin{eqnarray*}
&&\bm{B^*_M} =
\begin{bmatrix}
\begin{smallmatrix}
\bm{\pi}_{1,\gamma_1}[1]-1  &  \bm{\pi}_{1,\gamma_1}[2] & ... & \bm{\pi}_{1,\gamma_1}[M-1] & \bar{A}_{1,\gamma^*_1}\\
\bm{\pi}_{2,\gamma_2}[1]  &  \bm{\pi}_{2,\gamma_2}[2]-1 & ... & \bm{\pi}_{2,\gamma_2}[M-1] & \bar{A}_{2,\gamma^*_1}\\
...                         &  ...                      & ... & ...                       & ...\\
\bm{\pi}_{M,\gamma_M}[1]-1  &  \bm{\pi}_{M,\gamma_M}[2] & ... & \bm{\pi}_{M,\gamma_M}[M-1] & \bar{A}_{M,\gamma^*_1}
\end{smallmatrix}
\end{bmatrix}, \\
&&\bm{B^*} =
\begin{bmatrix}
\begin{smallmatrix}
\bm{\pi}_{1,\gamma_1}[1]-1  &  \bm{\pi}_{1,\gamma_1}[2] & ... & \bm{\pi}_{1,\gamma_1}[M-1] & \gamma^*_1\\
\bm{\pi}_{2,\gamma_2}[1]  &  \bm{\pi}_{2,\gamma_2}[2]-1 & ... & \bm{\pi}_{2,\gamma_2}[M-1] & \gamma^*_2\\
...                         &  ...                      & ... & ...                       & ...\\
\bm{\pi}_{M,\gamma_M}[1]-1  &  \bm{\pi}_{M,\gamma_M}[2] & ... & \bm{\pi}_{M,\gamma_M}[M-1] & \gamma^*_M
\end{smallmatrix}
\end{bmatrix}.
\end{eqnarray*}
\end{prop}

Finally, the performance of the relaxed greedy policy, i.e., the average sampled AoI, is given by the sum of $\bar{R}$ for the sensors in the set $\{n:\bar{\bm{h}}_n<\eta^*\}$:
\begin{eqnarray}\label{eq:finalperforamnce}
J(\hat{\mu}^\prime) = \frac{1}{\hat{d}(\eta^*)}\sum_{n=1}^{\sup\{n:\bar{\bm{h}}_n<\eta^*\}}\bar{R}_n,
\end{eqnarray}
where the constant $\frac{1}{\hat{d}(\eta^*)}$ is a correction term coming from \eqref{eq:dhat} as $\hat{d}(\eta^*)$ may not be exactly $1$.

\section{Lower and Upper Bounds to the Relaxed Greedy Policy}\label{sec:IV}
Before presenting the numerical and simulation results, we derive in this section lower and upper bounds to the average AoI sampled per slot of the relaxed greedy policy. Let us start from a symmetric setup where the error probabilities of sensors are the same.

\begin{thm}[lower and upper bounds in symmetric networks]\label{thm:symBounds}
In a symmetric network where the error probabilities of sensors $p_1=p_2=...=p_N=p$, the average sampled AoI of the relaxed greedy policy is bounded by $L^s_B\leq J(\hat{\mu}^\prime)\leq U^s_B$, where
\begin{eqnarray*}
&& L^s_B = \frac{1-p^{\zeta+1}}{1-p} + (1+2p-\zeta-p^{M-1})p^{\zeta+1},\\
&& U^s_B = \frac{1-p^{M+\zeta}}{1-p}-\zeta p^{\zeta +1},
\end{eqnarray*}
and $\zeta=\lfloor \frac{1-p^{M-1}}{1-p} \rfloor$, provided that the number of sensors in the network $N\geq 1 + (M-2)p^\zeta$.
\end{thm}

\begin{NewProof}
See Appendix \ref{sec:AppB}.
\end{NewProof}

Theorem \ref{thm:symBounds} bounds the performance of the relaxed greedy policy in a symmetric setting. In the general asymmetric setting, bounding the performance of the relaxed greedy policy turns out to be a nontrivial task. In the following, we shall present a universal lower bound to the average sampled AoI in \eqref{eq:metric}. In particular, by ``universal'', we mean that the bound is not tailored for the relaxed greedy policy, but for any possible sampling policies $\mu$ at the AP.

\begin{thm}[universal lower bound]\label{thm:LB}
A universal lower bound to \eqref{eq:metric} for any policy $\mu$ is given by
\begin{eqnarray}\label{eq:LB}
L_B = \sum_{n=1}^{N}\left[ \frac{(L^*\!-\! 1)p_n^{L^*}\!-\! L^*p_n^{L^*\!-\! 1}\! +\! 1}{q_n} \!+\!  q_n \omega^* L^*  p_n^{L^*\!-\!1}       \right],
\end{eqnarray}
where
\begin{eqnarray*}
&&\hspace{-0.5cm} L^* = \inf_L\left\{L:\sum_{n=1}^{N}(1-p_n^L)\geq 1     \right\}, \\
&&\hspace{-0.5cm} \omega^* = \inf_\omega\left\{\omega:\sum_{n=1}^{N}[1-\omega p_n^{L^*}- (1-\omega)p_n^{L^*-1}]\geq 1 \right\}.
\end{eqnarray*}
\end{thm}
\begin{NewProof}
See Appendix \ref{sec:AppC}.
\end{NewProof}

\begin{cor}
In the symmetric setting where $p_1=p_2=...=p_N=p$, the $L^*$ and $\omega^*$ in the lower bound \eqref{eq:LB} are given by
\begin{eqnarray*}
L^* = \left\lceil \log_p\left(1-\frac{1}{N}\right) \right\rceil, ~~
\omega^* = \frac{p^{L^*-1}+\frac{1}{N}-1}{p^{L^*-1}-p^{L^*}}.
\end{eqnarray*}
\end{cor}

\begin{NewProof}
See Appendix \ref{sec:AppC}.
\end{NewProof}

Next, we investigate random sampling as an upper bound to the greedy policy.

\begin{defi}[random sampling]
A random sampling policy instructs the AP to randomly sample the sensor, regardless of the belief state $\bm{\Pi}$.
\end{defi}

Proposition \ref{thm:randomsampling} below characterizes the average AoI received by the AP per time slot when operated with the random policy. We then prove in Theorem \ref{thm:randomBound} that random sampling is an upper bound to the relaxed greedy policy. The gap between random sampling and the relaxed greedy policy is quantified in Corollary \ref{thm:randomGAP}, considering the symmetric network setup.

\begin{prop}[Performance of random sampling]\label{thm:randomsampling}
With the random sampling policy, the average sampled AoI over the infinite horizon is given by
\begin{eqnarray}\label{eq:randomperformance1}
J_\text{random} =\frac{1}{N}\sum_{n=1}^{N}\frac{1-p_n^M}{1-p_n}.
\end{eqnarray}
In the symmetric setting where $p_1=p_2=...=p_N=p$, let $M\to\infty$, we have $J_\text{random}\to\frac{1}{1-p}$.
\end{prop}
\begin{NewProof}
The random sampling policy samples each sensor with probability $1/N$ in each slot. The average sampled AoI over the infinite horizon $J_\text{random}$ can then be written as $J_\text{random} =\frac{1}{N}\sum_{n=1}^{N}\bar{R}_n$, where $\bar{R}_n$ is the average sampled AoI from the $n$-th sensor.

For each sensor, the dynamic of the AoI is governed by the MC in Fig.~\ref{fig:MC}. The steady-state distribution of the MC, denoted by $\bm{h}_n$, is a left eigenvector (corresponding to eigenvalue one) of the stochastic matrix $\bm{\mathcal{T}}_n$, giving $\bm{h}_n \bm{\mathcal{T}}_n = \bm{h}_n$.
For the transition matrix $\bm{\mathcal{T}}_n$ in \eqref{eq:matrix}, solving $\bm{h}_n \bm{\mathcal{T}}_n = \bm{h}_n$ gives us
$\bm{h}_n =
[q_n,\allowbreak q_np_n,\allowbreak q_np_n^2,\allowbreak \cdots,\allowbreak q_n p_n^{M-2},\allowbreak p_n^{M-1}]$.
The expected AoI sampled from the $n$-th sensor is then $\bar{R}_n = \sum_{k=1}^{M-1} k q_n p_n^{k-1} + M p_n^{M-1} =\frac{1-p_n^M}{1-p_n}$.
Substituting $\bar{R}_n$ into $J_\text{random}$ gives us \eqref{eq:randomperformance1}.
\end{NewProof}

\begin{thm}\label{thm:randomBound}
The relaxed greedy policy is strictly better than random sampling: $J(\hat{\mu}^\prime)\leq J_{\text{random}}$.
\end{thm}

\begin{NewProof}
See Appendix \ref{sec:AppD}.
\end{NewProof}

\begin{cor}\label{thm:randomGAP}
In the symmetric settings where $p_1=p_2=...=p_N=p$, the gap between random sampling and the relaxed greedy policy is bounded by
\begin{eqnarray}
&& \frac{p^{M+\zeta}-p^M}{1-p}+\zeta p^{\zeta+1} \leq J_{\text{random}} - J(\hat{\mu}^\prime) \nonumber\\
&& \leq \frac{p^{\zeta+1}-p^M}{1-p}- (1+2p-\zeta-p^{M-1}) p^{\zeta+1}.
\end{eqnarray}
\end{cor}

\begin{NewProof}
Corollary \ref{thm:randomGAP} follows directly from Theorem \ref{thm:symBounds} and Proposition \ref{thm:randomsampling}.
\end{NewProof}

\section{Numerical and Simulation Results}\label{sec:VII}
This section presents the numerical and simulation results to compare the performance of the relaxed greedy policy and the greedy policy. Throughout this section, a large $M=100$ is set to avoid the impact of AoI truncation.
We shall consider four system settings to verify the approximation of the relaxed greedy policy to the greedy policy.
1) A symmetric network with equal error probabilities $\{p_n\}$. Such a setting can be mapped to scenarios where the sensors are deployed in a homogeneous environment, e.g., underwater (to test the water quality) or desert (to monitor dust storms).
2) An asymmetric network with deterministic error probabilities.
3) An asymmetric network with random error probabilities sampled from uniform distributions.
4) An asymmetric network with random error probabilities sampled from Gaussian distributions. Sensors with randomly-distributed error probabilities can be mapped to scenarios where there are a vast number of sensors. For example, when deployed in an area to monitor the state of a target, the qualities of the sensing channels, and hence the error probabilities, are determined by the distances from the sensors to the target. When the number of sensors is large, in general we can assume the error probabilities follow a Gaussian distribution.

\begin{figure}[t]
\centering
\begin{minipage}[t]{0.48\textwidth}
\centering
\includegraphics[width=6cm]{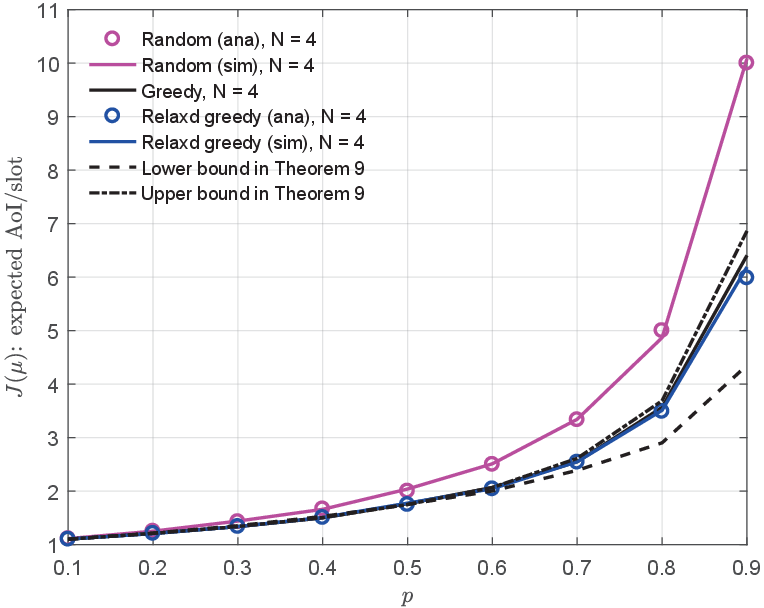}
\caption{The expected AoI sampled per slot in a symmetric network, where $N=4$, $M=100$.}
\label{fig:S1}
\end{minipage}
\begin{minipage}[t]{0.48\textwidth}
\centering
\includegraphics[width=5.7cm]{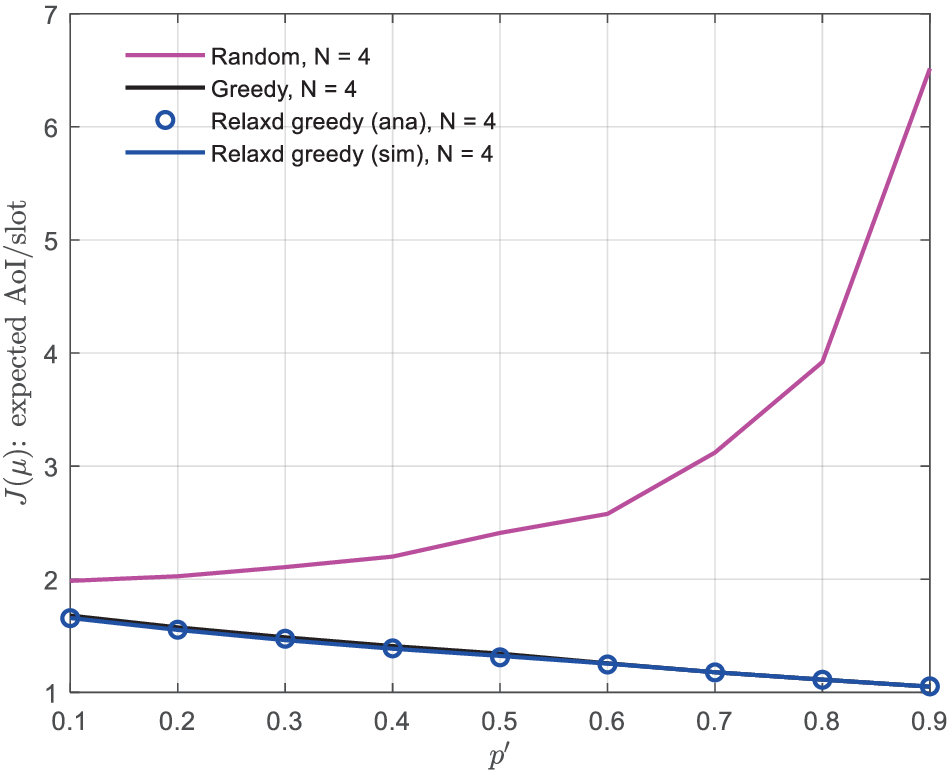}
\caption{The expected AoI sampled per slot in an asymmetric network with deterministic error probabilities.}
\label{fig:S2}
\end{minipage}
\end{figure}

Let us start from the symmetric network setting wherein the sensing channels of the $N$ sensors are equally good, i.e., the error probabilities $p_1=p_2=...=p_N=p$.
Fig. \ref{fig:S1} presents the expected AoI sampled per slot versus $p$ for the random policy, the greedy policy, and the relaxed greedy policy, respectively.
The lower and upper bounds of the relaxed greedy policy, as derived in Theorem \ref{thm:symBounds}, are also plotted.
The number of sensors in the network $N=4$.

As can be seen,
1) The expected AoI attained by the random policy is consistent with the analytical results. The performance gap between the random policy and the greedy policy increases as $p$ increases. When $p=0.9$, the AoI gap is about $3.8$.
2) For the relaxed greedy policy, our analytical result \eqref{eq:finalperforamnce} matches the simulation results as the two curves coincide with each other.
3) The relaxed greedy policy approximates the greedy policy very well. The gap between the two performance curves is negligible.

Next, we consider asymmetric networks where the qualities of the $N$ sensing channels vary and we have a set of error probabilities $\{p_n:n=1,2,...,N\}$. In particular, the set of error probabilities can be deterministic or random.

For the case of deterministic error probabilities, we follow two rules to set $\{p_n:n=1,2,...,N\}$: 1) the mean of the set of $p_n$ is fixed to $0.5$; 2) all the $p_n$ are equally spaced. As such, $p_n=0.5+(n-8)\frac{p^\prime}{N-1}$, $n=1,2,...,N$, where the constant $p^\prime$ is the span of $\{p_n\}$, i.e., $p^\prime =  \max p_n - \min p_n$.

The numerical and simulation results presented in Fig. \ref{fig:S2} yield the following observations:
\begin{enumerate}[\hspace{0.2pt}1)]
\item With the increase of $p^\prime$, the performance of the random policy deteriorates. This matches our intuition because a larger $p^\prime$ means more dispersed error probabilities, in which case the random sampling policy performs worse.
\item The approximation of the relaxed greedy policy to the greedy policy is accurate. The two curves coincide with each other.
\item With the increase of $p^\prime$, the performance of the greedy policy improves. This is because half of the sensing channels get better as $p^\prime$ increases, while the other half gets worse. With the greedy policy, AP samples the sensor with the minimum expected AoI in each slot. Therefore, the performance of the greedy policy is dominated by the better channels.\footnote{This implies that more sensors yield better performance in monitoring, but can also lead to a low sensor utilization rate since the performance is often dominated by the better channels. An interesting topic that is worth further study is the optimal number of sensors that strikes the best tradeoff between the average sampled AoI and the sensor utilization rate.} This explains the performance improvement of the greedy policy.
\end{enumerate}

For the case of random error probabilities, we assume the set of $\{p_n:n=1,2,...,N\}$ is sampled from a uniform distribution or a Gaussian distribution in an i.i.d. manner. For each policy, the expected-AoI performance is averaged over a large number of realizations of the distribution.

\begin{figure}[t]
  \centering
  \includegraphics[width=0.75\columnwidth]{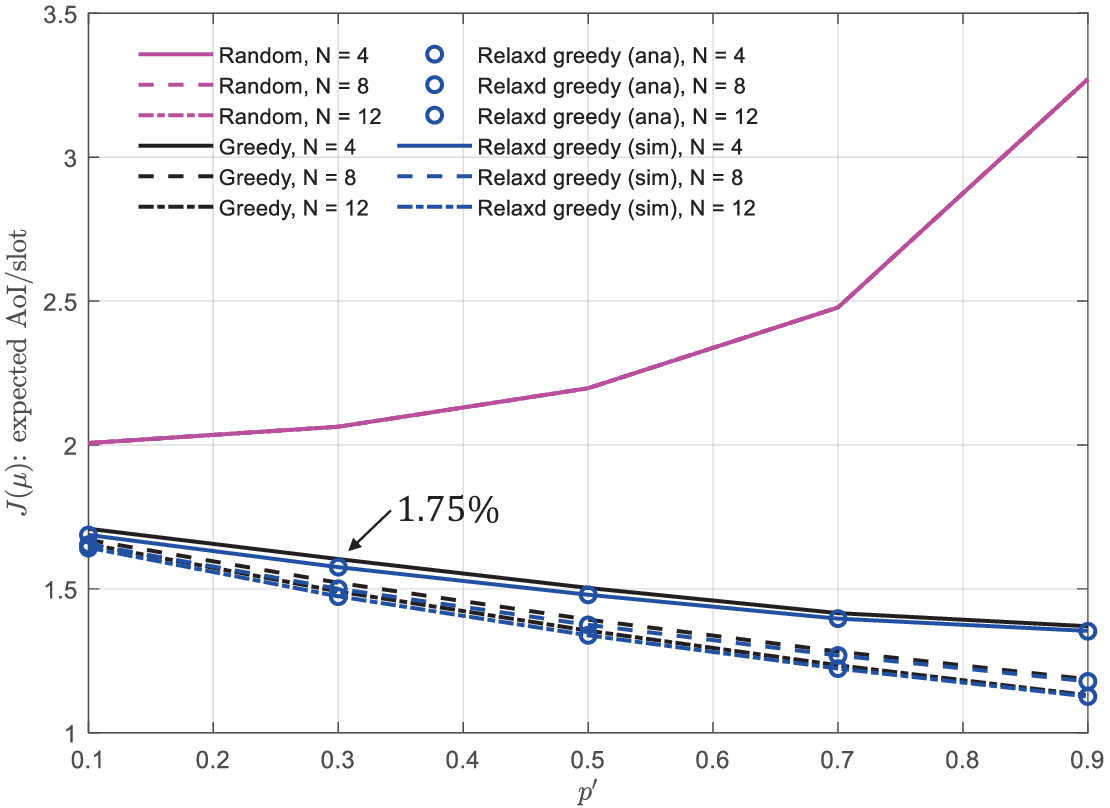}\\
  \caption{Performance  of  the  random  policy,  the  greedy  policy,  and  the  relaxed  greedy  policy  (numerical  and simulation) in an asymmetric network. The error probabilities are sampled from uniform distributions.}
\label{fig:S3}
\end{figure}

In Fig. \ref{fig:S3}, the set of error probabilities $\{p_n\}$ are sampled uniformly from the interval $[1/2\allowbreak-p^\prime/2,\allowbreak 1/2\allowbreak+p^\prime/2]$ (that is, $p^\prime$ is the width of the interval). The expected AoI per slot versus $p^\prime$ curves for different policies are plotted in Fig. \ref{fig:S3}, where the number of sensors $N=4$, $8$, and $12$.

A first observation is that the performance of the random policy is irrelevant to the number of sensors $N$. This can be understood from \eqref{eq:randomperformance1}. Specifically, when the error probabilities are sampled from a distribution $\varphi(x)$, we have
\begin{eqnarray}\label{eq:randomperformance3}
&&\hspace{-0.5cm} J_\text{random} = \mathbb{E}_{p_n\sim\varphi(x)}\left[\frac{1}{N}\sum_{n=1}^{N}\frac{1-p_n^M}{1-p_n}\right] \\
&&\hspace{-0.5cm}\overset{M\to\infty}{=}
\frac{1}{N}\sum_{n=1}^{N}\mathbb{E}_{p_n\sim\varphi(x)}\left[\frac{1}{1-p_n}\right]
\overset{(a)}{=}\mathbb{E}_{p_n\sim\varphi(x)}\left[\frac{1}{1-p_n}\right],\nonumber
\end{eqnarray}
where (a) follows because $\{p_n\}$ are sampled in an i.i.d. manner. Eq. \eqref{eq:randomperformance3} verified that $J_\text{random}$ is irrelevant to the number of sensors $N$. Furthermore, when $\varphi(x)$ is a uniform distribution in $[1/2\allowbreak-p^\prime/2,\allowbreak 1/2\allowbreak+p^\prime/2]$, we have
\begin{equation*}
J_\text{random}=\mathbb{E}_{p_n\sim\varphi(x)}\left[\frac{1}{1-p_n}\right] = \frac{1}{p^\prime}\ln\left(\frac{1+p^\prime}{1-p^\prime}\right).
\end{equation*}
This exactly matches the simulation results in Fig. \ref{fig:S3}.

The second observation from Fig. \ref{fig:S3} is that the performance of the greedy policy improves with the increase of the number of sensors. This can be easily understood since the  performance  of  the  greedy  policy  is  dominated  by  the  better  channels -- the more sensors there are, the higher the probability that a better channel (with lower error probability) is sampled from the uniform distribution. Moreover, the relaxed greedy policy approximates the greedy policy very well. The maximum gap between the two policies is only $1.75\%$, as shown in Fig. \ref{fig:S3}.

\begin{figure}[t]
  \centering
  \includegraphics[width=0.75\columnwidth]{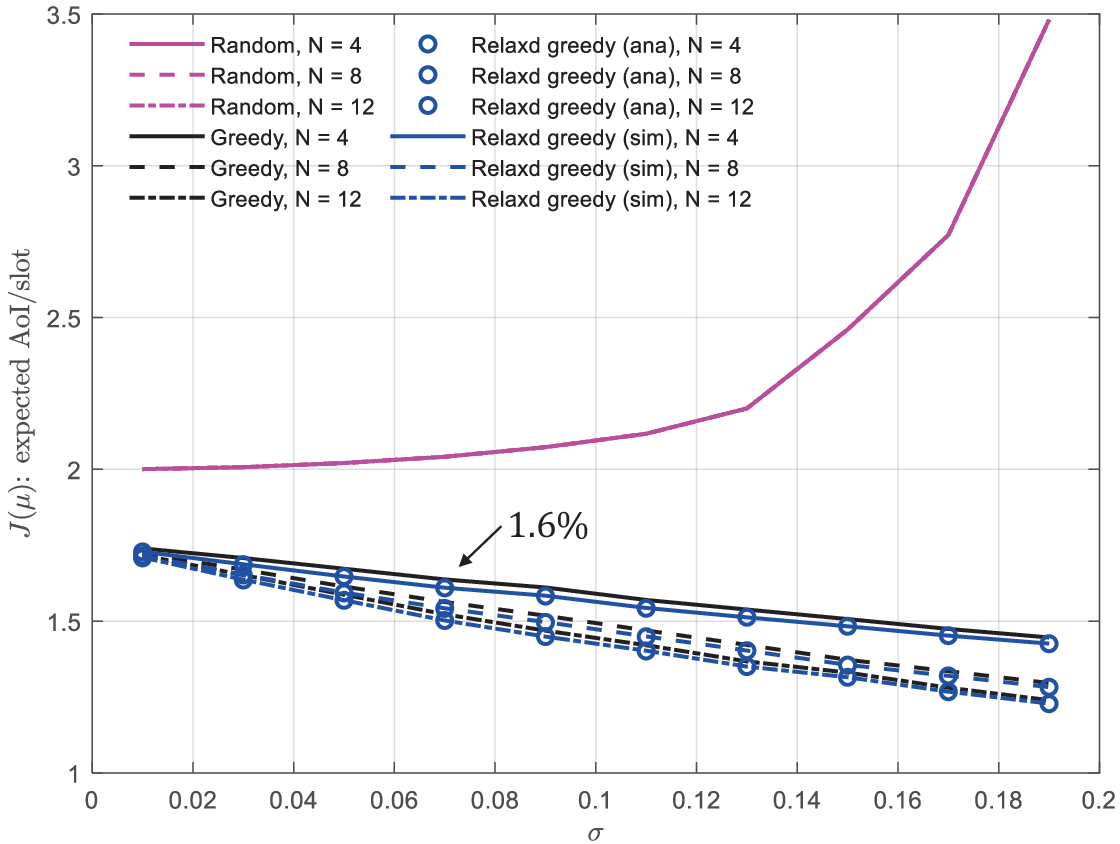}\\
  \caption{Performance  of  the  random  policy,  the  greedy  policy,  and  the  relaxed  greedy  policy  (numerical  and simulation) in an asymmetric network. The error probabilities are sampled from Gaussian distributions.}
\label{fig:S4}
\end{figure}

Finally, we repeat our simulations in Fig. \ref{fig:S3}, but sample the error probabilities $\{p_n\}$ from a Gaussian distribution with mean $1/2$ and standard deviation $\sigma$. The performance comparisons among different policies are shown in Fig. \ref{fig:S4}.
As per \eqref{eq:randomperformance3}, the performance of the random policy is irrelevant to the number of sensors $N$ if the error probabilities are sampled in an i.i.d. manner. This is verified in Fig. \ref{fig:S4}. For the relaxed greedy policy, our analytical results match the simulation results very well. Further, the maximum gap between the greedy and the relaxed greedy policies is only $1.6\%$ -- the relaxed greedy policy is thus an excellent approximation to the greedy policy.

\section{Conclusion}
Restless multi-armed bandit (RMAB) problems with partially observable arms are open problems due to the polynomial space (PSPACE) hardness of the partially observable Markov decision process (POMDP). This paper explored the greedy policy to solve this class of problems in the context of minimum-age scheduling.

In the minimum-age scheduling problem considered, an access point (AP) monitors the state of an object via a set of sensors. The ages of the sensed information (AoI) at different sensors vary and are unknown to the AP unless the AP samples them. Time is divided into slots. At any slot, the AP queries/samples one sensor to collect the most updated state information about the object. In particular, the sampling decision can only be made based on a sequence of past sampling decisions and AoI observations. The sampling process associated with each sensor is thus a POMDP with two possible actions ``sample'' and ``rest''. At any one time, only one action can be ``sample'' and the other actions are ``rest''. In general, the goal is to minimize the average sampled AoI over an infinite time horizon.

The greedy policy is the policy that attempts to minimize the average sampled AoI in the next immediate step rather than the average sampled AoI over an infinite time horizon. With the greedy policy, the AP compares the expected AoI that can be obtained from each sensor at a decision epoch and samples the sensor that yields the minimum expected AoI. Our goal in this paper was to analyze the averaged sampled AoI over an infinite horizon for the greedy policy.

The underpinning of our analysis was a relaxed greedy policy constructed to approximate the performance of the greedy policy. In each slot, the relaxed greedy policy instructs the AP to sample the sensors whose expected AoI is less than a constant, and the constant was chosen so that the AP samples one sensor per slot on average.

With the relaxed greedy policy, the RMAB is decoupled since the sampling process of each sensor is independent of the others. In particular, the decoupled problem of sampling a single sensor can be modeled as a POMDP with two possible actions -- sample or rest. Dissecting the inner structure of the POMDP gave us the average sampled AoI from the sensor, and finally, the performance of the relaxed greedy policy is a sum of the average sampled AoI from all sensors. Numerical and simulation results validated that the relaxed greedy policy is an excellent approximation to the greedy policy in terms of the expected AoI sampled over an infinite horizon.

\appendices

\section{}\label{sec:AppA}
\begin{figure}[t]
  \centering
  \includegraphics[width=0.65\columnwidth]{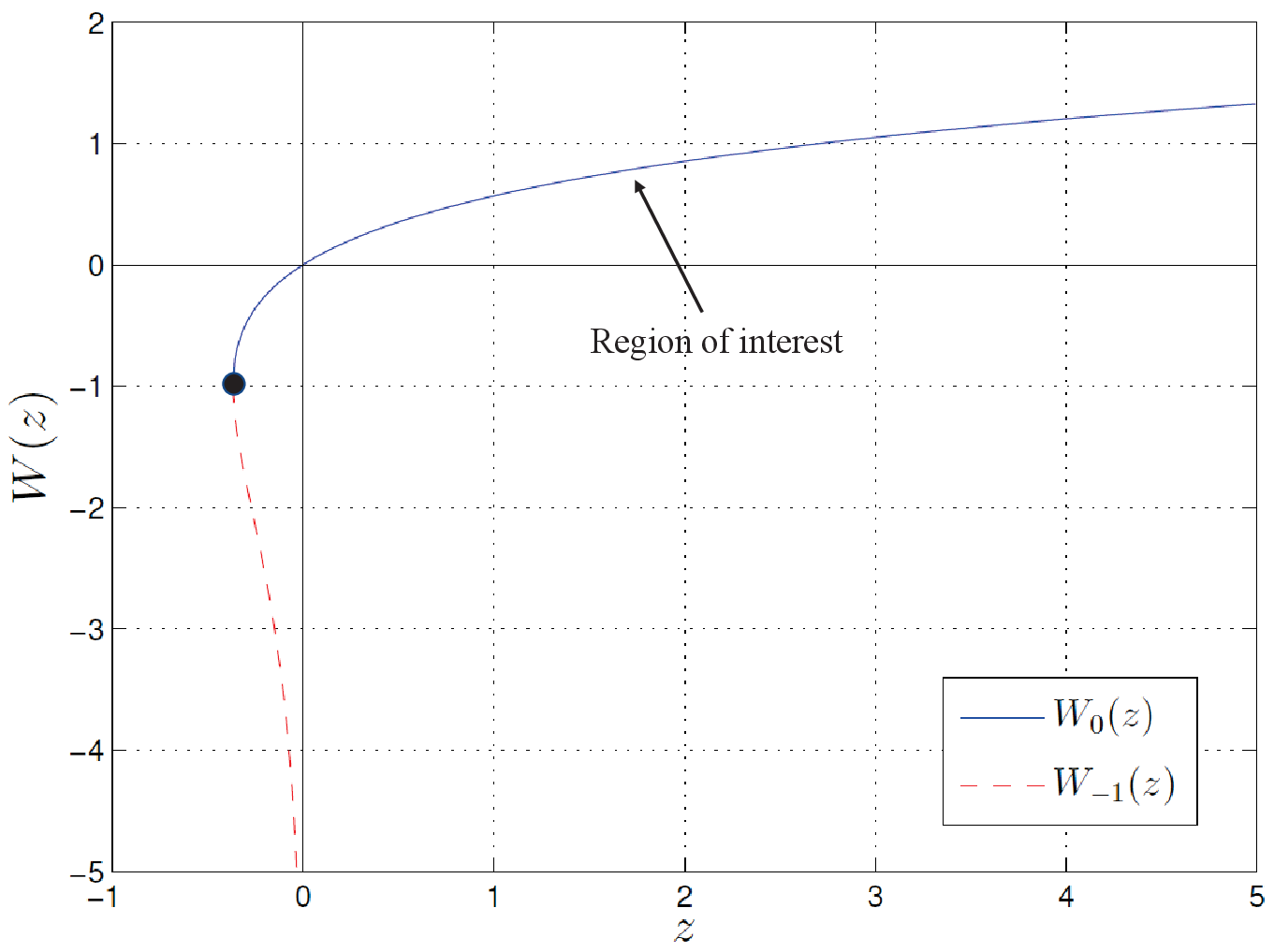}\\
  \caption{The two branches of a lambert W function for real input $z$.}
\label{fig:lambert}
\end{figure}
\begin{NewProof}
(sketch) For the POMDP associated with a single sensor, the evolution of the expected AoI follows Proposition \ref{thm:evolutions}. For each evolution path, $\gamma_k$ is given by \eqref{eq:thres}.
Let us first exclude the trivial case where $\eta$ is larger than the initial expected AoI of the $M$-th evolution branch, i.e.,
$\eta > \bar{A}_{M,1} = q + M p$.
This case is trivial because $\bar{A}_{M,1}$ is the largest expected AoI that can be obtained among all belief states.
If $\eta > \bar{A}_{M,1}$, we have $\eta > \bar{A}_{k,1}$, $\forall~k$, hence
$\gamma_k=1,~\forall~k$.

Next, we focus on $\eta:\bar{\bm{h}}< \eta \leq \bar{A}_{M,1}$. For the $k$-th evolution path,

1) 	Consider $\left\{k:\bar{A}_{k,1}=q+\min\{k+1,M\}p<\eta \right\}$. The sensor is sampled at the initial state in this case, thus $\gamma_k=1,~\forall~k:~\bar{A}_{k,1} <\eta$.

2) 	Consider  $\left\{k:\bar{A}_{k,1}\geq\eta,k\leq\frac{1}{1-p}\right\}$. As per Theorem \ref{thm:evolutions}, if $k\leq\frac{1}{1-p}$, the expected $\bar{A}_{k,1}$ first increases monotonically to $\bar{A}_{k,M-k}$ (the first phase), and then decreases monotonically to $\bar{A}_{k,M-1}=\bar{\bm{h}}$ (the second phase).Since $\bar{A}_{k,1}\geq\eta$, all the expected AoI in the first phase is larger than $\eta$, and the AP can only sample the sensor in the second phase. Let
\begin{eqnarray}\label{eq:AppC1}
\bar{A}_{k,x_1} = \frac{1-p^{x_1}}{1-p} - p^{x_1} x_1 + p^{x_1} M = \eta,
\end{eqnarray}
we have
\begin{eqnarray}\label{eq:AppC2}
x_1 =  \frac{1}{\ln p}W_0\left(\psi(\eta) p^{\psi(M)}\ln p\right) - \psi(M),
\end{eqnarray}
where $\psi(x)=\frac{1}{1-p}-x$ and $W_0(\star)$ is the principal branch of a Lambert W function. As shown in Fig.~\ref{fig:lambert}, a lambert W function $W(z)$ has two branches $W_0(z)$ and $W_{-1}(z)$ for real input $z$. Our region of interest is the principal branch.
Thus, $\gamma_k = \lceil x_1 \rceil,~\forall~k:~\bar{A}_{k,1}\geq\eta,k\leq\frac{1}{1-p}$.

3) Consider $\left\{k:\bar{A}_{k,1}\geq\eta,k>\frac{1}{1-p}\right\}$. As per Theorem \ref{thm:evolutions}, if $k>\frac{1}{1-p}$, the expected AoI $\bar{A}_{k,1}$ decreases monotonically to $\bar{A}_{k,M-1}=\bar{\bm{h}}$ in both the first phase and the second phase. Thus, the AP can sample the sensor in either phase.
If the AP sample the sensor in the first phase ($i+k\leq M$), we must have
$\eta > \bar{A}_{k,M-k-1} = \frac{1-p^{M-k-1}}{1-p} + kp^{M-k-1}$.

Let $\bar{A}_{k,x_2} = \frac{1-p^{x_2}}{1-p} + kp^{x_2} = \eta$,
we have $x_2 = \log_p \left( \frac{1-\eta(1-p)}{1-k(1-p)}  \right)$, and hence
\begin{eqnarray}\label{eq:AppC3}
\gamma_k= \lceil x_2 \rceil,~\forall~k:\bar{A}_{k,1}\geq\eta,k>\frac{1}{1-p},\bar{A}_{k,M-k-1}\leq \eta.
\end{eqnarray}

On the other hand, if the AP sample the sensor in the second phase ($i+k>M$), we have $\eta < \bar{A}_{k,M-k-1}$. In this case, the threshold can be derived as \eqref{eq:AppC1} and  \eqref{eq:AppC2}, giving
\begin{eqnarray*}
\gamma_k = \lceil x_1 \rceil,~\forall~k:\bar{A}_{k,1}\geq\eta,k>\frac{1}{1-p},\bar{A}_{k,M-k-1}> \eta.
\end{eqnarray*}

Then, we consider the case where $\eta$ is smaller than or equal to the steady-state expected AoI, i.e., $\eta<\bar{\bm{h}} = \frac{1-p^M}{1-p}$.
For the evolution branches whose initial states $\bar{A}_{k,1}\geq\eta$, the AP would never sample the sensor. As a result, $\gamma_k=\infty$.
If $\bar{A}_{k,1}<\eta$, the AP can sample the sensor in the first phase ($i+k\leq M$) according to Proposition \ref{thm:evolutions}. The threshold is given by \eqref{eq:AppC3}, i.e.,
\begin{eqnarray*}
\gamma_k= \left\lceil \log_p \left( \frac{1-\eta(1-p)}{1-k(1-p)}  \right) \right\rceil.
\end{eqnarray*}
In actuality, if $\eta$ is smaller than or equal to the steady-state expected AoI, the AP would never sample the sensor again if at some point the AP samples an AoI $k^\prime\in\{k^\prime:\bar{A}_{k^\prime,1} \geq \eta\}$ because $\gamma_{k^\prime}=\infty$.

Overall, we conclude that there are four kinds of thresholds for different evolution branches.
\begin{enumerate}
\item $\gamma=1$, where the AP samples the sensor in the initial state with expected AoI $\bar{A}_{k,1}$. This happens when $\{\eta > \bar{A}_{M,1}\}$ or $\left\{\bar{\bm{h}}< \eta \leq \bar{A}_{M,1}, \bar{A}_{k,1}<\eta \right\}$.
\item $\gamma = \lceil x_2 \rceil$, where the AP samples the sensor in the first phase of the evolution. This happens when $\{\bar{\bm{h}}<\allowbreak \eta \leq\allowbreak \bar{A}_{M,1},\allowbreak \bar{A}_{k,1}\geq\eta,k>\frac{1}{1-p},\allowbreak\bar{A}_{k,M-k-1}\leq \eta \}$ or $\{0\leq \eta\leq \bar{\bm{h}},\allowbreak \bar{A}_{k,1}<\eta \}$.
\item $\gamma = \lceil x_1 \rceil$, where the AP samples the sensor in the second phase of the evolution. This happens when $\{\bar{A}_{k,1}\geq\eta,k\leq\frac{1}{1-p} \}$ or $\{\bar{A}_{k,1}\geq\eta,k>\frac{1}{1-p},\bar{A}_{k,M-k-1}> \eta\}$.
\item $\gamma=\infty$, where the AP will never sample the sensor. This happens when $\{0\leq\eta\leq\bar{\bm{h}}, \bar{A}_{k,1}\geq\eta\}$.
\end{enumerate}

\end{NewProof}

\section{}\label{sec:AppB}
This appendix proves Theorem~\ref{thm:symBounds}.
Consider a randomly deployed wireless sensor network with $N$ sensors, as shown in Fig. \ref{fig:model}. The error probabilities of the sensors are $\bm{p}=\{p_n:n=1,2,...,N\}$. When operated with the relaxed greedy policy, a sensor is sampled by the AP if its expected AoI is smaller than a constant $\eta$. According to Theorem \ref{thm:thresholds}, for the $n$-th sensor, if $\eta$ is smaller than or equal to its steady-state average AoI, i.e., $0\leq \eta\leq \bar{\bm{h}}_n=\frac{1-p_n^M}{1-p_n}$, the AP would stop sampling the sensor at some point in time. Thus, we only need to focus on the sensors whose steady-state average AoI is smaller than $\eta$, i.e., $\eta>\bar{\bm{h}}_n$, and let $\eta=\bar{\bm{h}}_n +\delta_n$.

Consider a long sample path of $T$ slots. In the sampling process, we focus on a specific sensor with error probability $p_n$ and analyze the average AoI sample from it (for brevity, the subscript $n$ is omitted in the following). According to Lemma \ref{thm:prop2}, the belief states in the $k$-th evolution branch evolve as $\bm{\pi}_{k,i}=[q,qp,qp^2,...,qp^{i-1},0,...,0,p^i,0,...,0]$ for $i = 1,2,3,...,M-1$. Each evolution branch has a lifespan of $\gamma_k$ slots, i.e., the threshold $\gamma_k$ defined in Theorem \ref{thm:thresholds}.

For this sensor, we denote the average number of times it is sampled per slot by $d(p,\eta)$; the average AoI sampled per slot by $R(p,\eta)$; and the average AoI obtained per sample by $R^\prime(p,\eta)$ where
\begin{equation}\label{eq:A1}
R^\prime(p,\eta)=\frac{R(p,\eta)}{d(p,\eta)}.
\end{equation}
We further denote by $\lambda_z$ the number of the slots that an AoI of $z$ is sampled from the considered sensor in $T$ slots. Then, whenever the sensor is sampled, the probability that the sampled AoI is $z$ is defined as
\begin{equation}\label{eq:A2}
\phi[z]=\lim_{T\to\infty}\frac{\lambda_z}{Td(p,\eta)}.
\end{equation}
Given $\lambda_z$ and $\phi[z]$, we can write $R(p,\eta)$ and $d(p,\eta)$ as
\begin{equation}\label{eq:A4}
R(p,\eta)=\lim_{T\to\infty} \frac{1}{T}\sum_{z=1}^{M}\lambda_z,~~d(p,\eta)=\frac{1}{\sum_{z=1}^{M}\phi[z]\gamma_z},
\end{equation}
where \eqref{eq:A4} follows because $\sum_{z=1}^{M}\phi[z]d(p,\eta)T\gamma_z=T$.

Given the above definitions, we first prove a few useful results.

\begin{lem}\label{thm:1}
For the relaxed Greedy Policy, the average sampled AoI
\begin{equation}\label{eq:B1}
J(\hat{\mu}^\prime) = \frac{\sum_{i=1}^{N}R(p_i,\eta)}{\sum_{i=1}^{N}d(p_i,\eta)}.
\end{equation}
\end{lem}

\begin{NewProof}
Eq.~\eqref{eq:B1} follows directly from the definitions.
\end{NewProof}

\begin{lem}\label{thm:2}
For each sensor, if the thresholds of first $k$ evolution branches $\gamma_1=\gamma_2=\gamma_k=1$, we have $\phi[z]\geq qp^{z-1}$, $\forall z\in[1,k+1]$.
\end{lem}

\begin{NewProof}
We prove Lemma \ref{thm:2} by induction. Suppose a sensor is sampled in slot $t$ and let the sampled AoI be $i$. Then the sensor will be sampled again in slot $t+\gamma_i$, the corresponding belief state is $\bm{\pi}_{i,\gamma_i}$.

From Lemma \ref{thm:prop2}, we know that $\bm{\pi}_{i,\gamma_i}[1]=q$, $\forall i$. This means, whenever we sample, the probability that the sampled AoI is $1$ is $q$. That is,
\begin{equation*}
\phi[z=1]=q\geq q p^{z-1}.
\end{equation*}

Next, we assume $\phi[z]\geq q p^{z-1}$ holds for $z\in[1,k]$ and prove $\phi[k+1]\geq qp^k$.

Since $\gamma_k=1$, we have $\bm{\pi}_{k,\gamma_k}[i]=\bm{\pi}_{k,1}[i]=p$. This yields
\begin{equation*}
\phi[k+1]\geq p\phi[k]\geq p * qp^{k-1}\geq qp^k.
\end{equation*}

The lemma is proved.
\end{NewProof}

\begin{lem}\label{thm:3}
For any sensor, the average AoI obtained per sample $R^\prime(p,\eta)\leq \bar{\bm{h}}$.
\end{lem}

\begin{NewProof}
We prove Theorem \ref{thm:3} by contradiction. Suppose $R^\prime(p,\eta)> \bar{\bm{h}}$. We will prove that this leads to $\gamma_1=\gamma_2=...=\gamma_M=1$, and hence, $R^\prime(p,\eta) = \bar{\bm{h}}$.

As in Proposition 3, we define $\bar{A}_{k,i}=\sum_{j=1}^{M} j\bm{\pi}_{k,i}[j]$. At the beginning of the evolution branches, we have $i=1$ and $\bar{A}_{k,1}=1+kp$.

For the first branch $k=1$, it is easy to verify that $\bar{A}_{1,1}=1+p \leq \bar{\bm{h}} \leq \eta$. This implies that $\gamma_1=1$.
For any $k\in[1,M-2]$, we show that if $\gamma_1=\gamma_2=...=\gamma_k=1$, then $\gamma_{k+1}=1$.

From Lemma \ref{thm:2}, we know $\phi[i]\geq q p^{i-1}$, $\forall i\in [1,k+1]$. Thus, we can assume $\phi[i]= q p^{i-1}+\delta_i$, where $\delta_i\geq 0$. The average AoI obtained per sample can then be upper bounded by
\begin{equation}\label{eq:B2}
R^\prime(p,\eta)\leq \sum_{i=1}^{k} (1+ip)\phi[i] +\eta \left(1-\sum_{i=1}^{k}\phi[i] \right).
\end{equation}
This follows because 1) for the first $k$ branches, $\gamma_1=\gamma_2=...\gamma_k=1$, and hence, the average AoI is $1+ip$; 2) for the rest $M-k$ branches, the average AoI obtained per sample must be smaller than $\eta$.

The right-hand side (RHS) of \eqref{eq:B2} can be further refined as
\begin{eqnarray*}
\hspace{-0.2cm}&&\sum_{i=1}^{k} (1+ip)\phi[i] +\eta \left(1-\sum_{i=1}^{k}\phi[i] \right) \\
\hspace{-0.2cm}&&= \sum_{i=1}^{k} (1+ip)qp^{i-1} + p^i\eta + \sum_{i=1}^{k} (1+ip-\eta)\delta_i \\
\hspace{-0.2cm}&&\leq \sum_{i=1}^{k} (1+ip)qp^{i-1} +p^i \eta,
\end{eqnarray*}

We then have
\begin{equation*}
\frac{1-p^M}{1-p} = \bar{\bm{h}} < R^\prime(p,\eta) \leq \sum_{i=1}^{k} (1+ip)qp^{i-1} +p^i \eta.
\end{equation*}
This yields $\eta > 1+(1+k)p$, that is, $\gamma_{k+1}=1$. Overall, we have $\gamma_{1}=\gamma_{2}=...=\gamma_{M-1}=1$.

Finally, from Theorem \ref{thm:thresholds}, we have $\gamma_{M-1}=\gamma_{M}$, and hence, $\gamma_1=\gamma_2=...=\gamma_M=1$. This means $R^\prime(p,\eta) = \bar{\bm{h}}$, which contradicts the assumption $R^\prime(p,\eta)> \bar{\bm{h}}$. As a result, $R^\prime(p,\eta)\leq \bar{\bm{h}}$.
\end{NewProof}

\begin{lem}\label{thm:4}
For any sensor, the thresholds of the evolution branches $\gamma_1=\gamma_2=...=\gamma_\zeta=1$, where $\zeta=\lfloor\frac{1-p^{M-1}}{1-p} \rfloor$.
\end{lem}

\begin{NewProof}
Lemma \ref{thm:4} follows from Theorem \ref{thm:thresholds}. In particular, $\gamma_\zeta=1$ if and only if
\begin{equation*}
\bar{A}_{1,1} \leq \bar{A}_{2,1} \leq ... \leq \bar{A}_{\zeta,1} = 1 +\zeta p \leq \bar{\bm{h}} = \frac{1-p^M}{1-p}.
\end{equation*}
Thus, $\zeta$ is at most $\lfloor\frac{1-p^{M-1}}{1-p} \rfloor$.
\end{NewProof}

Next, we bound the performance of the relaxed greedy policy under a symmetric network setup, wherein the error probabilities of all sensors are equal, i.e., $p_1 =...p_N=p$.

\begin{prop}\label{thm:5}
In a symmetric network where $p_1=p_2=...p_N=p$, the constant $\eta^*$ of the relaxed greedy policy is $\eta^*=\bar{\bm{h}}+\varepsilon$, where $\varepsilon\to 0$, if the number of sensors $N\geq 1+(M-2) p^\zeta$, where $\zeta=\lfloor\frac{1-p^{M-1}}{1-p} \rfloor$.
\end{prop}

\begin{NewProof}
In order for a sensor to be persistently sampled by the AP, its steady-state average AoI $\bar{\bm{h}}$ must be smaller than $\eta^*$ (see Theorem \ref{thm:thresholds}). Thus, in the symmetric case, we have $\eta^*>\bar{\bm{h}}$ because no sensor would be sampled otherwise. Recall from Section V-B that $\eta^*$ is the largest $\eta$ so that the AP samples one sensor per slot on average. This implies that we only need to prove the following argument: when $\eta^*=\bar{\bm{h}}+\varepsilon$ and $\varepsilon\to 0$, the average number of sensors being sampled per slot $\sum_{i=1}^N d(p,\eta)=Nd(p,\eta)\geq 1$.

For each node, we have from \eqref{eq:A4} and that
\begin{eqnarray}\label{eq:C1}
d(p,\eta)\hspace{-0.3cm}&=&\hspace{-0.3cm}\frac{1}{\sum_{k=1}^{M}\phi[k]\gamma_k}
\overset{(a)}{=} \frac{1}{\sum_{k=1}^{\zeta}\phi[k]\! +\! \sum_{k=\zeta+1}^{M}\phi[k]\gamma_k} \nonumber\\
\hspace{-0.3cm}&\geq&\hspace{-0.3cm} \frac{1}{\sum_{k=1}^{\zeta}\phi[k]+\sum_{k=\zeta+1}^{M}\phi[k](M-1)},
\end{eqnarray}
where (a) follows from Lemma \ref{thm:4}.

Then, from Lemma \ref{thm:2}, we know $\phi[z]\geq qp^{z-1}$, $\forall z\in [1,k+1]$, if $\gamma_1=\gamma_2=...=\gamma_k=1$. Let $\phi[z]=qp^{z-1}+\delta_z$, the RHS of \eqref{eq:C1} can be written as
\begin{eqnarray}\label{eq:C2}
&& \frac{1}{\sum_{k=1}^{\zeta}(qp^{k-1}\!+\!\delta_k)\!+\!\left(1\!-\!\sum_{k=1}^{\zeta}(qp^{k-1}\!+\!\delta_k)\right) (M\!-\!1)} \nonumber \\
&& =\frac{1}{\sum_{k=1}^{\zeta}qp^{k-1} + p^\zeta(M-1)+\sum_{k=1}^\zeta (2-M)\delta_k} \nonumber \\
&& \geq \frac{1}{1+p^\zeta(M-2)}.
\end{eqnarray}
Combining \eqref{eq:C1} and \eqref{eq:C2} gives us
\begin{eqnarray*}
\sum_{i=1}^N=Nd(p,\eta)\geq \frac{N}{1+p^\zeta(M-2)}.
\end{eqnarray*}
As a result, $\sum_{i=1}^N d(p,\eta) \geq 1$ holds if $N\geq 1 + (M-2)p^\zeta$.

\end{NewProof}

In a symmetric network, the average sampled AoI of the relaxed greedy policy
\begin{equation*}
J(\hat{\mu}^\prime)=\frac{\sum_{i=1}^{N}R(p,\eta)}{\sum_{i=1}^{N}d(p,\eta)}=\frac{R(p,\eta)}{d(p,\eta)}=R^\prime(p,\eta).
\end{equation*}
Thus, we can focus on the sampling process of a single sensor to bound the average AoI obtained per sample $R^\prime(p,\eta)$.

To derive the lower bound, we have
\begin{eqnarray*}
\hspace{-0.5cm}&& J(\hat{\mu}^\prime) = R^\prime(p,\eta) \geq \sum_{i=1}^{\zeta}\phi[i]\bar{A}_{1,i} + \phi[\zeta+1]\bar{\bm{h}} + \sum_{i=\zeta+2}^{M}\phi[i]\bar{\bm{h}} \\
\hspace{-0.5cm}&&\overset{(a)}{=} \sum_{i=1}^{\zeta}(qp^{i-1}+\delta_i)(1+ip) + (qp^\zeta+\delta_{\zeta+1})\frac{1-p^M}{1-p} +\\
\hspace{-0.5cm}&&\hspace{0.5cm} \left(1-\sum_{i=1}^{\zeta+1}(qp^{i-1}+\delta_i) \right) \frac{1-p^M}{1-p} \\
\hspace{-0.5cm}&&= \sum_{i=1}^{\zeta} qp^{i-1} (1+ip) + qp^\zeta \frac{1-p^M}{1-p} +
\left(1-\sum_{i=1}^{\zeta+1}qp^{i-1} \right)* \\
\hspace{-0.5cm}&&\hspace{0.5cm} \frac{1-p^M}{1-p} + \sum_{i=1}^{\zeta}\delta_i\left(1+ip-\frac{1-p^M}{1-p} \right)\\
\hspace{-0.5cm}&&\overset{(b)}{\geq} \sum_{i=1}^{\zeta} qp^{i-1} (1+ip) + qp^\zeta \frac{1-p^M}{1-p} +
\left(1-\sum_{i=1}^{\zeta+1}qp^{i-1} \right) * \\
\hspace{-0.5cm}&&\hspace{0.5cm}\frac{1-p^M}{1-p} + \sum_{i=2}^{\zeta}\delta_i\left(1+2p-\frac{1-p^M}{1-p} \right)\\
\hspace{-0.5cm}&&\overset{(c)}{\geq} \sum_{i=1}^{\zeta} qp^{i-1} (1+ip) + qp^\zeta \frac{1-p^M}{1-p} +
\left(1-\sum_{i=1}^{\zeta+1}qp^{i-1} \right) * \\
\hspace{-0.5cm}&&\hspace{0.5cm} \frac{1-p^M}{1-p} + \left(1+2p-\frac{1-p^M}{1-p} \right)p^{\zeta+1}\\
\hspace{-0.5cm}&&= \frac{1-p^{\zeta+1}}{1-p} + (1+2p-\zeta-p^{M-1})p^{\zeta+1},
\end{eqnarray*}
where (a) follows by setting $\phi[z]=qp^{z-1}+\delta_z$, as in \eqref{eq:C2}; (b) follows because $\delta_1=0$; and (c) follows from $\sum_{i=1}^{\zeta+1}(qp^{i-1}+\delta_i)\leq 1$, from which we have $\sum_{i=1}^{\zeta+1}\delta_i\leq p^{\zeta+1}$. As a result,
\begin{eqnarray*}
\sum_{i=2}^\zeta\delta_i\leq p^{\zeta+1}-\delta_{\zeta+1}\leq p^{\zeta+1}.
\end{eqnarray*}

To derive the upper bound, we have
\begin{eqnarray*}
\hspace{-0.5cm}&& J(\hat{\mu}^\prime) = R^\prime(p,\eta) \leq \sum_{i=1}^{\zeta}\phi[i]\bar{A}_{1,i} + \sum_{i=\zeta+1}^{M}\phi[i]\eta^* \\
\hspace{-0.5cm}&& = \sum_{i=1}^{\zeta}(qp^{i-1}+\delta_i)(1+ip) + (1-\sum_{i=1}^{\zeta}(qp^{i-1}+\delta_i))\eta^* \\
\hspace{-0.5cm}&& \leq \sum_{i=1}^{\zeta}(qp^{i-1}+\delta_i)(1+ip) + p^\zeta \eta^* = \frac{1-p^{M+\zeta}}{1-p} - \zeta p^{\zeta+1}.
\end{eqnarray*}

\section{}\label{sec:AppC}
This appendix proves Theorem \ref{thm:LB}. The corollary of Theorem \ref{thm:LB} is proved in part D of this appendix.
To derive the lower bound, we shall 1) propose a fictitious sampling policy called the proactive transmission policy; 2) analyze the performance of the proactive transmission policy; 3) prove that the performance of the proactive transmission policy is a universal lower bound to \eqref{eq:metric} if a constraint is imposed.
\subsection{The proactive Transmission Policy}
For the system model considered in this paper, the sensors report the sensed data to the AP in a passive way. As described in Section \ref{sec:II}, the AP determines which sensor to transmit in the next slot and triggers the transmission by the beacon message. To construct the lower bound, we consider the sampling problem from a different angle: we let the sensors proactively transmit their sensed data if their data is fresh enough \cite{IFDMA}.

\begin{defi}[proactive transmission policy]\label{defi:proactive}
With the proactive transmission policy, a sensor 1) transmits the sensed information to the AP with probability $1$ if the AoI is less than $L$; 2) transmits its sensing data to the AP with probability $\omega$ if the AoI equals $L$; 3) does not transmit if the AoI is larger than $L$.
\end{defi}

In the following, we first derive the performance of the proactive transmission policy and then prove it is a lower bound to \eqref{eq:metric} under some constraints.

\subsection{The Sampling Process of a Single Sensor}
With the proactive transmission policy, the transmissions among different sensors are decoupled because a sensor decides to transmit or not based only on its own AoI. Therefore, we can simply consider the transmission process of a single sensor.

As shown in Fig. \ref{fig:MC}, for a single sensor, the transitions of AoI form a Markov Chain. Let us consider one evolution trajectory of the AoI that starts from the $a^0=1$., i.e., a sequence of AoI of this sensor in consecutive slots. In particular, we take the state $1$ (i.e., AoI equals $1$) as a reference point.

Over time, the evolution trajectory goes back to state $1$ repeatedly. Let us call a subsequence of AoI that starts from state $1$, ends with the state $1$, and with no state $1$ in between a ``cycle''. Then, the duration of a cycle(in terms of the number of slots), denoted by $\mathbb{I}$, is a random variable. Specifically, we have
\begin{eqnarray}
\textup{Pr}(\mathbb{I}=I) = p^{I-1} q,~I = 1,2,3,...
\end{eqnarray}
where $I$ is a realization of the random variable $\mathbb{I}$.

Given a cycle of length $I$, we have the following results:
\begin{enumerate}
\item If $I\leq M$, the AoI in this cycle increases from $1$ to $I$ monotonically (increased by $1$ in every slot); If $I>M$, the AoI in this cycle increases from $1$ to $M$ and then remains in state $M$ until the cycle terminates.
\item The sensor transmits its sensed data to the AP only when the AoI is smaller than or equal to $L$ ($L\leq M$). Thus, the number of transmissions in this cycle is
\begin{eqnarray}
c_s(I) = \begin{cases}
I, & \text{if}~I\leq L-1, \\
L-1+\omega, & \text{if}~I\geq L.
\end{cases}
\end{eqnarray}
Note that the sensor transmits with probability $\omega$ if the AoI equals $L$.
\item The sum of AoI transmitted in this cycle is
\begin{eqnarray}
G_s(I) = \begin{cases}
\frac{(I+1)I}{2}, & \text{if}~I\leq L-1, \\
\frac{(L-1)L}{2}+L\omega, & \text{if}~I\geq L.
\end{cases}
\end{eqnarray}
Note that if $I>L$, there is no transmission, and the AoI $I$ is not added to the above tally.
\end{enumerate}

Suppose there are $Q$ cycles in the evolution trajectory ($Q\to\infty$). Then, we have
\begin{enumerate}
\item The total number of slots in the trajectory is
\begin{eqnarray}
T = \sum_{I=1}^{\infty}\textup{Pr}(\mathbb{I}=I)QI=\frac{Q}{q}.
\end{eqnarray}
\item The total number of transmissions in the trajectory is
\begin{eqnarray}
&&\hspace{-1cm} c_\textup{all} = \sum_{I=1}^{\infty}\textup{Pr}(\mathbb{I}=I)Qc_s(I) \\
&&\hspace{-1cm} =\sum_{I=1}^{L-1}p^{I-1}qQI+\sum_{I=L}^{\infty}p^{I-1}qQ(L-1+\omega) \nonumber\\
&&\hspace{-1cm} = Q \left\{\frac{(L-1)p^L-Lp^{L-1}+1}{q}+(L-1+\omega)p^{L-1}\right\}.\nonumber
\end{eqnarray}
\item The total transmitted AoI in the trajectory is
\begin{eqnarray}
&&\hspace{-1cm} G_\textup{all}=\sum^{\infty}_{I=1}\textup{Pr}(\mathbb{I}=I)QG_s(I) \\
&&\hspace{-1cm} = \sum^{L-1}_{I=1}p^{i-1}qQ\frac{(I+1)I}{2}+\sum_{I=L}^{\infty}p^{I-1}qQ\left[\frac{(L-1)L}{2}+L\omega \right]\nonumber\\
&&\hspace{-1cm} = Q \left\{\frac{(L-1)p^L-Lp^{L-1}+1}{q^2}+\omega L p^{L-1}\right\}.\nonumber
\end{eqnarray}
\end{enumerate}

As a result, for a single sensor, the number of transmissions per slot is
\begin{eqnarray*}
\frac{c_\textup{all}}{T} &&\hspace{-0.5cm}= (L-1) p^L-Lp^{L-1}+1+(L-1+\omega)qp^{L-1} \nonumber\\
&&\hspace{-0.5cm} =1-\omega p^L-(1-\omega) p^{L-1},
\end{eqnarray*}
and the average AoI transmitted per slot is
\begin{eqnarray}\label{eq:AppA1}
\frac{G_\textup{all}}{T}=\frac{(L-1)p^L-Lp^{L-1}+1}{q} + \omega L q p^{L-1}.
\end{eqnarray}

\subsection{A Lower Bound to \eqref{eq:metric}}
Given \eqref{eq:AppA1}, the performance of the proactive transmission policy (i.e., the average AoI received by the AP per slot) is simply the sum of the average AoI transmitted from each sensor to the AP. Thus,
\begin{eqnarray}\label{eq:AppA2}
J_\textup{pro}(L,\omega)\!=\!\sum_{n=1}^{N}\!\left[ \frac{(L\!-\!1)p_n^L\!-\!Lp_n^{L\!-\!1}\!+\!1}{q_n} \!+\! \omega L q_n p_n^{L\!-\!1} \right].
\end{eqnarray}
Meanwhile, the number of transmissions to the AP per slot is
\begin{eqnarray*}
\sum_{n=1}^{N}\left[ 1-\omega p_n^L-(1-\omega) p_n^{L-1} \right].
\end{eqnarray*}

In order to make \eqref{eq:AppA1} a lower bound to \eqref{eq:metric}, we have to tune $L$ and $\omega$ so that there is on average one transmission per slot. That is, $L$ and $\omega$ are subject to the following constraint
\begin{eqnarray}\label{eq:AppA4}
L,\omega = \left\{L,\omega:  \sum_{n=1}^{N}\left[ 1-\omega p_n^L-(1-\omega) p_n^{L-1} \right]=1     \right\}.
\end{eqnarray}

Let
\begin{eqnarray*}
f(L,\omega) = \sum_{n=1}^{N}\left[ 1-\omega p_n^L-(1-\omega) p_n^{L-1} \right].
\end{eqnarray*}
It is easy to show that $f(L,\omega)$ is a monotonically increasing function of $L$ and $\omega$.

Denote the pair of $L$ and $\omega$ that satisfy \eqref{eq:AppA4} by $L^*$ and $\omega^*$. Let $\omega=1$, we have
\begin{eqnarray}
L^*=\inf_L\left\{L:\sum_{n=1}^{N}(1-p^L_n)\geq 1    \right\}.
\end{eqnarray}
Substituting $L^*$ into \eqref{eq:AppA4} yields
\begin{eqnarray*}
\omega^* = \inf_\omega\left\{\omega:\sum_{n=1}^{N}[1-\omega p_n^{L^*}- (1-\omega)p_n^{L^*-1}]\geq 1     \right\}.
\end{eqnarray*}

For such $L^*$ and $\omega^*$, the $J_\textup{pro}(L^*,\omega^*)$ in \eqref{eq:AppA2} is the lower bound to \eqref{eq:metric}. To see why this is true, let us consider the AoI transmitted to the AP in consecutive $T$ slots ($T\to\infty$). As per definition \ref{defi:proactive}, the proactive transmission policy can transmit 1) zero packets to the AP in a slot, in which case the AoIs of all sensors are larger than or equal to $L$; 2) one and only one packet to the AP in a slot, in which case the AoI of the transmitter is the smallest among all the sensors; 3) more than one packets to the AP in a slot, in which case the AoI of all the transmitters is less than or equal to $L$.

In a slot $t$, $t=1,2,...,T$, let the set of sensors that transmit in the slot be $\mathcal{S}^t$. Based on the above three cases, we can partition the $T$ slots into three subsets. Subset $\varphi_0$ consists of slots in which no sensor transmits, i.e., $\varphi_0=\{t:|\mathcal{S}^t |=0\}$; Subset $\varphi_1$ consists of slots in which only one sensor transmits, i.e., $\varphi_1=\{t:|\mathcal{S}^t |=1\}$; Subset $\varphi_2$ consists of slots in which more than one sensor transmit, i.e., $\varphi_2=\{t:|\mathcal{S}^t |\geq 2\}$.

As per \eqref{eq:AppA4}, the sum of the cardinality of $\mathcal{S}^t$ is $T$ because there is on average one transmission per slot. That is,
\begin{eqnarray}\label{eq:AppA5}
\sum_{t=1}^{T}|\mathcal{S}^t| = |\varphi_1| + \sum_{t\in \varphi_2} |\mathcal{S}^t| = T.
\end{eqnarray}
The AoI collected by the AP in $T$ slots is
\begin{eqnarray}
J^*_\textup{pro} =\sum_{t\in \varphi_1} a^t_{n^*} + \sum_{t\in \varphi_2}\sum_{i\in \mathcal{S}^t} a^t_i
\end{eqnarray}
where $n^*$ is the only element in $\mathcal{S}^t$ for $t\in \varphi_1$.

We now compare the proactive transmission policy with any sampling policy $\mu$ in \eqref{eq:metric}. Suppose the policy $\mu$  samples the $n^t$-th sensor in slot $t+1$ and obtains an AoI of $a_{n^t}^t$.

1) For each slot in the subset $\varphi_1$, there is only one sensor transmitting, the AoI of which is the minimum among all other sensors. This gives us
\begin{eqnarray}\label{eq:AppA6}
\sum_{t\in \varphi_1} a^t_{n^*} \leq \sum_{t\in \varphi_1}a_{n^t}^t.
\end{eqnarray}

2) For each slot in the subset $\varphi_2$, there is more than one sensor transmitting, but the AoI of these transmitting sensors is less than or equal to $L$, according to the definition. In addition, from \eqref{eq:AppA5}, the number of transmissions in the subset $\varphi_2$ is
\begin{eqnarray}
\sum_{t\in \varphi_2} |\mathcal{S}^t|  = T - |\varphi_1|.
\end{eqnarray}

This gives us
\begin{eqnarray}\label{eq:AppA7}
\sum_{t\in \varphi_2}\sum_{i\in \mathcal{S}^t} a^t_i &&\hspace{-0.5cm}= \sum_{t\in \varphi_2} a^t_{i^*_t} + \sum_{t\in \varphi_2}\sum_{i\in \mathcal{S}^t\setminus i^*} a^t_i \nonumber\\
&&\hspace{-0.5cm} \leq \sum_{t\in \varphi_2} a^t_{n^t} + \sum_{t\in \varphi_0} a^t_{n^t},
\end{eqnarray}
where $i_t^*=\arg\min_{i\in \mathcal{S}^t} a_i^t$ is the sensor with the minimum AoI in slot $t\in \varphi_2$. The inequality follows because 1) the number of AoI summed together on both the LHS and RHS is $T-|\varphi_1|$; 2) the first term on the LHS is smaller than or equal to the first term on the RHS because the AoI of the $i_t^*$-th sensor is the minimum for a slot $t\in \varphi_2$. Policy $\mu$ can never sample an AoI smaller than the AoI of the $i^*$-th sensor; 3) the second term on the LHS is smaller than or equal to the second term on the RHS because $a_i^t\leq L$ while $a_{n^t}^t\geq L$.

Combining \eqref{eq:AppA6} and \eqref{eq:AppA7}, we have
\begin{eqnarray*}
J^*_\textup{pro} &&\hspace{-0.5cm} =\sum_{t\in \varphi_1} a^t_{n^*} + \sum_{t\in \varphi_2}\sum_{i\in \mathcal{S}^t} a^t_i \\
&&\hspace{-0.5cm} \leq  \sum_{t\in \varphi_1} a^t_{n^t} + \sum_{t\in \varphi_2} a^t_{n^t} + \sum_{t\in \varphi_0} a^t_{n^t} = J(\mu).
\end{eqnarray*}
The performance of the proactive transmission policy is a lower bound to $J(\mu)$.

\subsection{The Lower Bound in the Symmetric Setting}\label{sec:AppA_E}
In the symmetric setting where $p_1=p_2=...=p_N=p$, the constraint \eqref{eq:AppA4} can be written as
\begin{eqnarray*}
N\left[ 1-\omega p_n^L-(1-\omega) p_n^{L-1} \right]=1
\end{eqnarray*}
After some manipulation, we have
\begin{eqnarray}\label{eq:AppA8}
(p^{L-1}-p^L)\omega + 1 - p^{L-1} - \frac{1}{N} = 0.
\end{eqnarray}

Let
\begin{eqnarray*}
f(L,\omega) = (p^{L-1}-p^L)\omega + 1 - p^{L-1} - \frac{1}{N},
\end{eqnarray*}
where $L=1,2,3,...$ and $\omega\in(0,1]$; $f(L,w)$ is a monotonically increasing function of $L$ and $\omega$. Thus, given any $L$, we have
\begin{equation*}
1\!-\!p^{L\!-\!1}\!-\!\frac{1}{N} \!=\! f(L,0)\!<\!f(L,\omega)  \!\leq\! f(L,1)\! = \!1\!-\!p^{L}\!-\!\frac{1}{N},
\end{equation*}
and $f(1,0)=-1/N$. Thus, there exists a unique $L^*$ such that
\begin{equation}
\begin{cases}
1 - p^{L^*-1} - \frac{1}{N} < 0,\\
1-p^{L^*}-\frac{1}{N} \geq 0.
\end{cases}
\end{equation}

This gives us
\begin{eqnarray*}
L^* = \left\lceil \log_p\left(1-\frac{1}{N}\right) \right\rceil.
\end{eqnarray*}

Substituting $L^*$ into \eqref{eq:AppA8} yields
\begin{eqnarray*}
\omega^* = \frac{p^{L^*-1}+\frac{1}{N}-1}{p^{L^*-1}-p^{L^*}}.
\end{eqnarray*}

\section{}\label{sec:AppD}
This appendix proves Theorem \ref{thm:randomBound}. Without loss of generality, we assume the error probabilities of the $N$ sensors are sorted in ascending order, i.e., $p_1\leq p_2\leq ...\leq p_N$. According to Lemma \ref{thm:1}, we have
\begin{eqnarray*}
&&J(\hat{\mu}^\prime) = \frac{\sum_{i=1}^N R(p_i,\eta)}{\sum_{i=1}^N d(p_i,\eta)}
= \frac{\sum_{i=1}^N R^\prime(p_i,\eta)d(p_i,\eta)}{\sum_{i=1}^N d(p_i,\eta)} \\
&&= \sum_{i=1}^N R^\prime(p_i,\eta)\frac{d(p_i,\eta)}{\sum_{i=1}^N d(p_i,\eta)}
= \sum_{i=1}^N R^\prime(p_i,\eta) d^\prime(p_i,\eta)\\
&&\leq \sum_{i=1}^N \bar{\bm{h}}_i d^\prime(p_i,\eta)
\end{eqnarray*}
where $d^\prime(p_i,\eta)$ is defined as $d^\prime(p_i,\eta)=\frac{d(p_i,\eta)}{\sum_{i=1}^N d(p_i,\eta)}$.

Since $p_1\leq p_2\leq ... \leq p_N$, we have $\bar{\bm{h}}_1\leq \bar{\bm{h}}_2\leq ...\bar{\bm{h}}_N$ and $d^\prime(p_1,\eta)\geq d^\prime(p_2,\eta)\geq ...\geq d^\prime(p_N,\eta)$. For any $i,j\in[1,N]$, $(\bar{\bm{h}}_i-\bar{\bm{h}}_j)(d^\prime(p_i,\eta)-d^\prime(p_j,\eta))\leq 0$ suggests that
\begin{eqnarray*}
\bar{\bm{h}}_i d^\prime(p_i,\eta) + \bar{\bm{h}}_j d^\prime(p_j,\eta)
\leq \bar{\bm{h}}_i d^\prime(p_j,\eta) + \bar{\bm{h}}_j d^\prime(p_i,\eta).
\end{eqnarray*}

Therefore,
\begin{eqnarray*}
&& 2J(\hat{\mu}^\prime) = 2N\sum_{i=1}^N \bar{\bm{h}}_i d^\prime(p_i,\eta) \\
&& = N\sum_{i=1}^N \bar{\bm{h}}_i d^\prime(p_i,\eta) + N\sum_{j=1}^N \bar{\bm{h}}_j d^\prime(p_j,\eta) \\
&& = \sum_{i=1}^N\sum_{j=1}^N (\bar{\bm{h}}_i d^\prime(p_i,\eta)+\bar{\bm{h}}_j d^\prime(p_j,\eta)) \\
&& \leq \sum_{i=1}^N\sum_{j=1}^N (\bar{\bm{h}}_i d^\prime(p_j,\eta)+\bar{\bm{h}}_j d^\prime(p_i,\eta)) \\
&& = \sum_{i=1}^N\sum_{j=1}^N \bar{\bm{h}}_i d^\prime(p_j,\eta) + \sum_{i=1}^N\sum_{j=1}^N \bar{\bm{h}}_j d^\prime(p_i,\eta) \\
&& = \sum_{i=1}^N \bar{\bm{h}}_i + \sum_{j=1}^N \bar{\bm{h}}_j
= 2\sum_{i=1}^N \bar{\bm{h}}_i
= 2 J_{\text{random}}.
\end{eqnarray*}
Thus, we have $J(\hat{\mu}^\prime)\leq J_{\text{random}}$.

\bibliographystyle{IEEEtran}
\bibliography{References}

\end{document}